\providecommand{\tabularnewline}{\\}

\input{aipcheck}

\def\tod{\mathop{{\to}}\limits}
\def\simd{\mathop{{\sim}}\limits}

\documentclass[
    ,final            
  ,numberedheadings 
  ]
  {aipproc}

\layoutstyle{6x9}

\usepackage{color}

\begin{document}

\title{Equilibrium and out of equilibrium phase transitions in systems with
long range interactions and in 2D flows}


\author{Freddy Bouchet}{address={Institut Non Lineaire de Nice, INLN, UMR 6618,  CNRS, UNSA, 1361 route des lucioles, 06 560 Valbonne - Sophia Antipolis, France}
}

\author{Julien Barr\'e}{address={Universit\'e de Nice-Sophia Antipolis,
Laboratoire J.~A.~Dieudonn\'e, Parc Valrose, 06108 Nice Cedex 02, France }
}

\author{Antoine Venaille}{address={Coriolis-LEGI, 21 avenue des Martyrs, 38000 Grenoble, France} 
}

\begin{abstract}
  In self-gravitating stars, two dimensional or geophysical flows and
  in plasmas, long range interactions imply a lack of additivity for
  the energy; as a consequence, the usual thermodynamic limit is not
  appropriate. However, by contrast with many claims, the equilibrium
  statistical mechanics of such systems is a well understood subject.
  In this proceeding, we explain briefly the classical approach to
  equilibrium and non equilibrium statistical mechanics for these
  systems, starting from first principles. We emphasize recent and
  new results, mainly a classification of equilibrium phase
  transitions, new unobserved equilibrium phase transition, and out of
  equilibrium phase transitions. We briefly discuss what we consider
  as challenges in this field.
\end{abstract}

\maketitle


\section{Introduction}

In a large number of physical systems, any single particle feels a
potential dominated by interactions with far away particles: this is
our definition of long range interactions. In a system with algebraic
decay of the inter-particle potential
$V\left(r\right)\simd_{r\rightarrow\infty}r^{\alpha}$, this occurs
when $\alpha$ is less than the dimension of the system (these
interactions are sometimes called ''non-integrable''). Then the energy
is not additive, as the interaction of any subpart of the system with
the whole is not negligible with respect to the internal energy of
this given part.

Self gravitating stars, after the discovery of negative specific heat
in \cite{LyndenBell:1968_MNRAS}, have played a very important
historical role, by emphasizing the peculiarities in the statistical
mechanics of systems with long range interactions. Besides
astrophysical self gravitating systems
\cite{Klinko_Miller_2002_PRE,Youngkins_Miller_2000_PRE,Stahl_Kiessling_Schindler_1995,Miller_Youngkins_1998_PRL,Gross_Votyakov_2000_EPJB,Chavanis_Rieutord_2003_AA,Heggie_Hut_2003,Ispolatov_Cohen_houches_2002,De_Vega_Sanchez_2002_NuclPhysB,Hertel_Thirring_1971_AnnPhys},
the main physical examples of non-additive systems with long range
interactions are two-dimensional or geophysical fluid dynamics
\cite{CagliotiLMP:1995_CMP_II(Inequivalence),Kiessling_Lebowitz_1997_LMP,SommeriaRobert:1991_JFM_meca_Stat,Miller:1990_PRL_Meca_Stat,Chavanis_houches_2002,Bouchet_Sommeria:2002_JFM}
and a large class of plasma effective models
\cite{Elkens_Escande_2002,Kiessling_Neukirch_2003_PNAS,Firpo_Elskens_PRL,free_electron_laser_2004_PRE}.
Spin systems \cite{BarreMukamelRuffo:2001_PRL_BEC} and toy models with
long range interactions
\cite{DRAW:2002_Houches,Antoni_Ruffo_1995PRE,Antoni_Ruffo_Torcini_2002PRE}
have also been widely studied. The links between these different
subjects
have been emphasized recently \cite{DRAW:2002_Houches}.\\

In these systems, the most prominent and interesting physical
phenomenon is the self organization of the particles, or of the
velocity field. This leads to coherent clouds of particles in plasma
physics, to galaxies and globular clusters in astrophysics and to large
scale jets and vortices in two dimensional or geophysical flows. Given
the large number of particles or of degrees of freedom, it is tempting
to adopt a statistical approach in order to describe these phenomena.
The statistical description of such a self organization, both at the
levels of equilibrium situations and relaxation towards equilibrium
(kinetic theories), is a classical, long studied field. One of the aim
of this proceeding is to insist on the vitality of this old subject
and to stress new advances and remaining issues. By contrast, the out
of equilibrium statistical mechanics of such phenomena is still in its
infancy, and few studies have been devoted to it. We emphasize the
importance of such studies for real applications, as most plasma and
geophysical physical phenomena are out of equilibrium. We also
describe some recent very suggestive results.

Both equilibrium and out of equilibrium phase transitions play a key
role in our understanding of physics, because they separate regions of
parameter space with qualitatively different behaviors. Very
naturally, a large part of our studies will be devoted to phase
transitions. We will especially stress the peculiar association of
phase transitions with negative specific heat and statistical ensemble
inequivalence in systems with long range interactions. We also insist
on recently observed out of equilibrium phase transtions, in the
context of two dimensional flows.  Finally, we describe our personal
guesses for what may be the challenges and interesting issues in the
field of systems with long range interactions. We hope this could open
new discussions, although we are conscious that such guesses are
necessarily biased by personal prejudices. We actually sincerely hope
that future researches will be much richer than what we describe. The
article is organized in three main sections: equilibrium, relaxation
to equilibrium and kinetic theories, non equilibrium stationary states.

\paragraph{Equilibrium}

Long range interacting systems are known to display peculiar
thermodynamic behaviors. As additivity is often seen as a cornerstone
of usual statistical mechanics and thermodynamics, it is sometimes
written in textbooks or articles that {}``statistical mechanics or
thermodynamics do not apply to systems with long range interactions''.
In this paper, we argue on the contrary that usual tools and ideas of
statistical mechanics do apply to such systems, both at equilibrium
and out of equilibrium.  However, reviewing a variety of recent works,
we will show that a careful application of these tools reveals truly
unusual and fascinating behaviors, absent from the world of short
range interacting systems.

After a brief introduction on the unusual negative specific heat and
other peculiar thermodynamical phenomena, we discuss the usual
assumptions of equilibrium statistical mechanics and their
interpretation in systems with long range interactions. Based on the
assumption of equal probability of any configuration with a given
energy, we then explain why the Boltzmann-Gibbs entropy actually
measures the probability to observe a given distribution function.
This relies on our ability to prove large deviations results for such
systems. The result of this analysis is that microcanonical and
canonical ensembles of systems with long range interactions are
described by two dual variational problems. We explain why such
variational problems lead to possible generic ensemble inequivalence,
and to a richer zoology of phase transitions than in usual systems. A
natural question then arises: do we know all possible behaviors
stemming from long range interactions, and, if not, what are the
possible phenomenologies? We answer this question by discussing a
\emph{classification} of all microcanonical and canonical phase
transitions, in long range interacting systems, with emphasis on
situations of ensemble inequivalence~\cite{Bouchet_Barre:2005_JSP}.
Very interestingly many possible phase transitions and situations of
ensemble inequivalence have not been observed yet. We then describe,
for two dimensional flows, the first observation of appearance of
ensemble inequivalence associated to bicritical and azeotropy phase
transitions.\\

\paragraph{Kinetic theories and relaxation toward equilibrium}

Because systems with long range interactions relax very slowly towards
equilibrium, or because they can be forced by external field, the
study of out of equilibrium situations is physically essential. During
the past century, there have been many attempts to find a general
formalism for out of equilibrium statistical mechanics, which would
give the equivalent of the Gibbs picture for out of equilibrium states.
Unfortunately, as recognized by most of the statistical mechanics
community, until now any such attempt failed. This is mainly due to
the fact that our knowledge of out of equilibrium situations can not
be parameterized by a small number of macroscopic quantities, playing
the same role as dynamical invariants for the equilibrium theory.
Then out of equilibrium statistical mechanics must be addressed by
a case by case careful examination of dynamics, using some appropriate
probabilistic description.

For relaxation to equilibrium of Hamiltonian systems with long range
interactions, standard tools have been developed, mainly kinetic
theory. In the introductory paragraph, we briefly explain the basic
ideas of kinetic theories. In the following, we first stress the role
of a Vlasov description for small time, and then the role of
Lenard-Balescu equation (also called collisional Boltzmann equation in
the context of self-gravitating systems) for larger time. We also
briefly review the recent application of Lynden-Bell equilibrium
statistical mechanics for the Vlasov equation to simple one
dimensional models. We also discuss new recent results for the kinetic
theory of such systems. The first is the generic existence, for the
one particle stochastic process, of anomalous diffusion and of long
relaxation times. We guess that the implications of such a result for
the validity of the kinetic approach has not been well appreciated up
to now. The second class of results deals with the time of validity
for the Vlasov approximation and with the typical time needed to
observe relaxation towards equilibrium. One of the most striking
result is that the Lenard Balescu operator vanishes identically for
one dimensional systems. This explains the existence of anomalous
scaling laws for the relaxation towards equilibrium
in models like the HMF model. \\

\paragraph{Non equilibrium stationary states (NESS)}

Another class of out of equilibrium problems is the study of systems
with long range interactions subjected to small non-Hamiltonian forces
and to weak dissipation. Such a framework is actually the most
relevant one for many physical applications. We will emphasize its
interest for geophysical flows, like for instance simplified models of
ocean currents.

The average balance between forcing and dissipation usually leads to
statistically stationary states, the properties of which may be
studied experimentally, numerically and theoretically. As there is no
detailed balance, the system is maintained out of equilibrium. The
fluxes of the Hamiltonian conserved quantities then become essential
physical variables.

We show in this last section, that this leads, in the context of two
dimensional flows, to very interesting out of equilibrium phase
transitions. We believe that the study of the statistical mechanics of
such non equilibrium stationary states and phase transitions, in other
systems with long range interactions, is one of the main challenges in
this field.

\section{Equilibrium statistical mechanics of systems with long range interactions
\label{sec:Equilibrium}}

\subsection{Peculiarities of thermodynamics of systems with long range interactions}

\label{sec:Generalite}

For systems with long range interactions, the most intriguing
thermodynamical property is the generic occurrence of statistical
ensemble inequivalence and negative specific heat. Such possibilities
have first been recognized and studied in the context of self
gravitating systems
\cite{LyndenBell:1968_MNRAS,Hertel_Thirring_1971_CMP,Hertel_Thirring_1971_AnnPhys}.
Afterwards, ensemble inequivalence and negative specific heat have been
observed or predicted in a number of different physical systems: two
dimensional turbulence
\cite{CagliotiLMP:1995_CMP_II(Inequivalence),Smith_ONeil:1990_phys_fluid,EllisHavenTurkington:2002_Nonlinearity_Stability},
plasma physics
\cite{Smith_ONeil:1990_phys_fluid,Kiessling_Neukirch_2003_PNAS}, spin
systems or toy models
\cite{BarreMukamelRuffo:2001_PRL_BEC,Antoni_Ruffo_Torcini_2002PRE}, or
self gravitating systems in situations different from the simple
initial case
\cite{Chavanis_Rieutord_2003_AA,Miller_Youngkins_1998_PRL,Youngkins_Miller_2000_PRE,Stahl_Kiessling_Schindler_1995,Gross_Votyakov_2000_EPJB,Tatekawa_Bouchet_DR:2005_PRE,Ispolatov_Cohen_houches_2002}.
A detailed description of each of these cases is provided in
\cite{Bouchet_Barre:2005_JSP}.

\begin{figure}
\includegraphics[height=0.2\textheight,keepaspectratio]{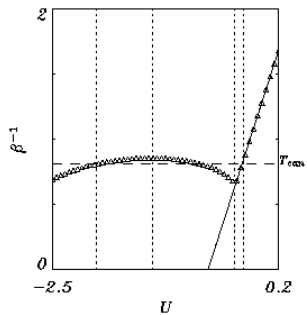}

\caption{Caloric curve (temperature $T=\beta^{-1}$ as a function of
  the energy $U$) for the SGR model, for $\epsilon=10^{-2}$. The
  energies corresponding to the dashed vertical lines are referred to,
  from left to right, as $U_{low}$, $U_{top}$, $U_{c}$, and
  $U_{high}$. The decreasing temperature between $U_{top}$ and $U_{c}$
  characterizes a range of negative specific heat and thus of ensemble
  inequivalence. At $U_{c}$, there is a microcanonical second order
  phase transition associated with the canonical first order phase
  transition (please see the text for a detailed explanation). Such a
  behavior is linked to the existence of tricritical points in both
  statistical ensembles (see figure \ref{fig:Tricritique_SGR}).  A
  classification of all possible routes to ensemble inequivalence is
  briefly described in section \ref{sec:Classification} or in
  \cite{Bouchet_Barre:2005_JSP}.}

\label{fig:Caloric_SGR}
\end{figure}

To motivate the following development, let us give an example where
ensemble inequivalence appears in an unusual way. We discuss the
equilibrium properties of the Self Gravitating Ring (SGR) model, a toy
model for self gravitating systems. Whereas we mainly present here its
equilibrium properties, we stress that this system is very interesting
also from a dynamical point of view, as it shows a number of out of
equilibrium quasi-stationary states
\cite{Iguchi_SGR_2005_PRE,Sota_SGR_2001_PRE}.

The Hamiltonian of the SGR model is:
$H=\frac{1}{2}\sum_{i=1}^{N}p_{i}^{2}-\frac{1}{2}\sum_{i,j=1}^{N}\frac{1}{\sqrt{2}}\frac{1}{\sqrt{1-cos(\theta_{i}-\theta_{j})+\epsilon}}$.
Particles are constrained on a ring ($0\leq\theta_{i}\leq2\pi$).  The
angles $\theta_{i}$ are conjugate to the momenta $p_{i}$. $\epsilon$
is a small scale softening of the gravitational interaction. We study
the phase transitions of this system and how they evolve when
$\epsilon$ is varied. Please see \cite{Tatekawa_Bouchet_DR:2005_PRE}
for a detailed discussion.

Figure \ref{fig:Caloric_SGR} shows the caloric curve $T(U)$ for
$\epsilon=10^{-2}$, where $T$ is the temperature and $U$ is the
energy. For $U_{top}<U<U_{c}$ the temperature is decreasing. The
specific heat $C=dU/dT$ is thus negative in this area, showing that
statistical ensembles are not equivalent (in the canonical ensemble,
the specific heat is always positive). The horizontal dashed line
is the Maxwell construction, which links microcanonical and canonical
ensembles. From it one sees that in the canonical ensemble, when $\beta$
is varied, there is a first order phase transition characterized by
an energy jump between the values $U_{low}$ and $U_{high}$. This
is a common feature in case of ensemble inequivalence.

\begin{figure}

\caption{Phase diagrams for the SGR model. Left: for each value of $\epsilon$,
the critical energies where a microcanonical phase transition occurs
are represented by black points or circles. When the softening parameter
is varied, at the tricritical point $\epsilon=\epsilon_{T}^{\mu}$,
first order phase transitions (black points) change to second order
phase transitions (circles). Right: the bottom dashed line and black
line represent respectively $U_{low}$ (see fig. \ref{fig:Caloric_SGR})
and $U_{high}$. The difference $U_{high}-U_{low}$ is the energy
jump associated to a first order phase transition in the canonical
ensemble. This jump decreases to zero at canonical tricritical point
$\epsilon_{T}^{c}$ where the canonical first order phase transition
changes to a canonical second order phase transition. The tricritical
points have different energies and $\epsilon$ values in both ensembles.}

\label{fig:Tricritique_SGR}

\begin{tabular}{cc}
\includegraphics[height=0.2\textheight]{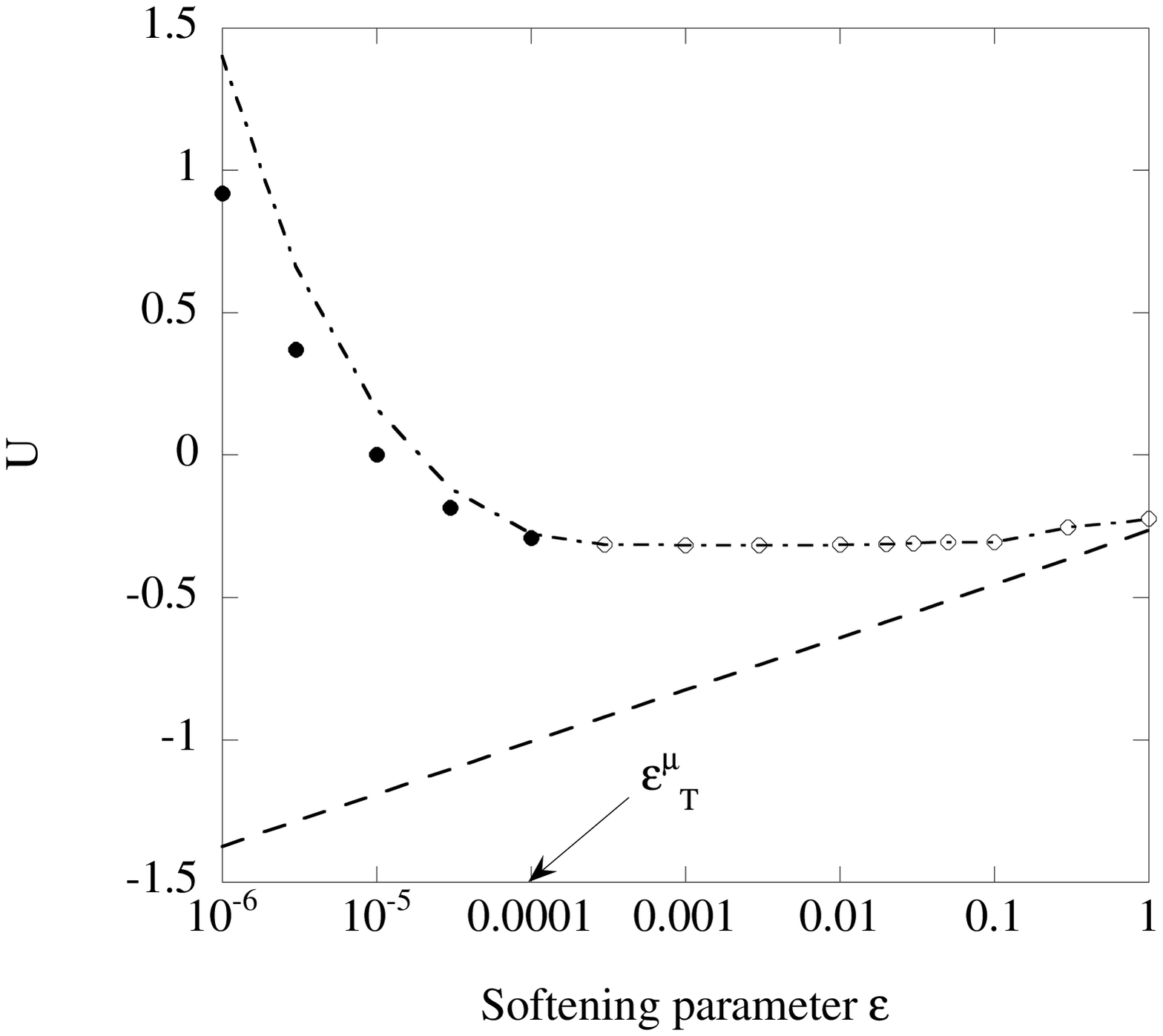}&
\includegraphics[height=0.2\textheight]{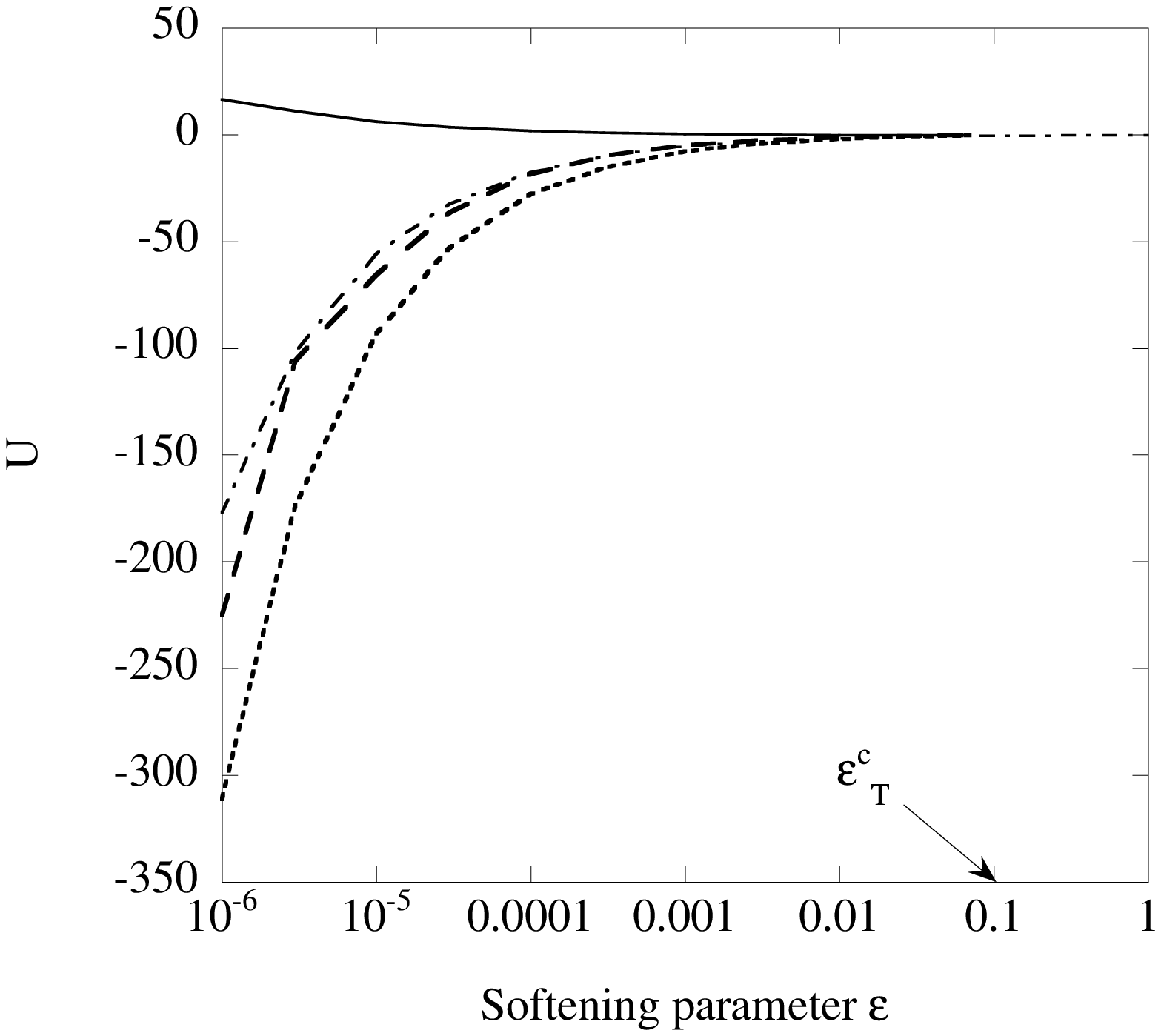} \tabularnewline
\end{tabular}
\end{figure}

What is less common is the concomitant existence of a second order
phase transition in the microcanonical ensemble, at the energy $U_{c}$.
At this point, the temperature $T$ is continuous, whereas its derivative
is discontinuous as it is clear from the curve. In figure \ref{fig:Tricritique_SGR},
we show that this type of ensemble inequivalence, with the coexistence
of a first order canonical first order phase transition and of a microcanonical
second order phase transition, is linked to the existence of a tricritical
point in both ensembles. \\

The SGR model displays one possible route to ensemble inequivalence,
out of several others. However, there are some constraints on the
possible phenomenologies. For instance, at a second order phase
transition, the negative specific heat jump must be positive. By
contrast, the temperature jumps at a discontinuity associated with a
first order microcanonical phase transition must be negative (this
means that when energy is increased the system has a negative
temperature jump).  Summarizing all these constraints yields a
\emph{classification} \cite{Bouchet_Barre:2005_JSP} of all possible
ensemble inequivalences and their links with phase transitions. To
prepare the discussion of this classification, we recall now the main
hypothesis and definitions of statistical equilibrium, in the context
of systems with long range interactions.

\subsubsection{Additivity extensivity and thermodynamic limit}

\label{sec:Additivity}

When studying statistical mechanics of non additive systems, the first
problem one has to deal with is the inadequacy of the thermodynamic
limit ($N\to\infty$, $V\to\infty$ -$V$ is the volume-, with $N/V$ kept
constant). Indeed, what is physically important in order to understand
the behavior of large systems, is not really to study the large $N$
limit, but rather to obtain properties that do not depend much on $N$
for large $N$ (the equivalent of intensive variables). For short range
interacting systems, this is achieved through the thermodynamic limit;
for non additive systems, the scaling limit to be considered is
different and depends on the problem. Let us consider $N$ particles
which dynamics is described by the Hamiltonian \begin{equation}
  \centering
  H_{N}=\frac{1}{2}\sum_{i=1}^{N}p_{i}^{2}+\frac{c}{2}\sum_{i,j=1}^{N}V(x_{i}-x_{j})~,\label{eq:hamiltonian}\end{equation}
where $c$ is a coupling constant. The thermodynamic limit in this case
amounts to send $N$ and the volume to infinity, keeping density and
$c$ constant. If $V(x)$ decreases fast enough so that interactions for
a particle come mainly from the first neighbors, then increasing $N$
at constant density has almost no effect on the bulk, and physical
properties are almost independent of $N$: the thermodynamic limit is
appropriate. This is wrong of course if the potential for a particle
is dominated by the influence of far away particles. The appropriate
scaling in this case may be as follows: fixed volume,
$c\propto1/N^{2}$, and $N\to\infty$ (others equivalent combinations
are possible, as the one given below for self gravitating particles).


The best known example of such a special scaling concerns self gravitating
stars, for which the ration $M/R$ is usually kept constant, where
$M$ is the total mass and $R$ is the system's radius (thermodynamic
limit would be $M/R^{3}$ constant). Another toy example is given
and studied for instance in \cite{Tamarit_Anteneodo_2000_PhRvL}.
This type of scaling is also the relevant one for point vortices in
two dimensional and geophysical turbulence, where the total volume
and total vorticity have to be kept fixed, but divided in smaller
and smaller units. Let us note for completeness that in some cases,
the thermodynamic limit is appropriate in presence of long range interactions,
for instance when some screening is involved~\cite{Grousson_Tarjus_Viot_2000_PhRvE};
we shall exclude these cases in the following.

According to the above discussion, let us rewrite Eq.~\ref{eq:hamiltonian}
using the convenient scaling $c=\pm1/N$:

\begin{equation}
H_{N}=\frac{1}{2}\sum_{i=1}^{N}p_{i}^{2}\pm\frac{1}{2N}\sum_{i,j=1}^{N}V(x_{i}-x_{j})\label{eq:HN}\end{equation}

This classical scaling of the coupling parameter is called the Kac's
prescription (\cite{Kac_1963_JMP}) or sometimes the mean field scaling
(see for instance \cite{Messer_Spohn_1982_JSP}). Within this scaling,
taking the limit $N\rightarrow\infty$ with all other parameters fixed
(fixed volume for instance), the sum over $i$ and $j$ is clearly
of order $N^{2}$, and the energy per particle $H_{N}/N$ is intensive.
This scaling is also the relevant one in order to obtain the collisionless
Boltzmann equation, for the dynamics, in the large $N$ limit. We
will use Eq.~\ref{eq:HN} in the following.

\subsubsection{The microcanonical and canonical ensembles}

We suppose that the energy $E$ of our system is known, and consider
the microcanonical ensemble. In this statistical ensemble all phase
space configurations with energy $E$ have the same probability; the
associated microcanonical measure is then

\[
\mu_{N}=\frac{1}{\Omega_{N}\left(E\right)}\prod_{i=1}^{N}dx_{i}dp_{i}\delta\left(H_{N}\left(\left\{ x_{i},p_{i}\right\} \right)-E\right),\]

where $\Omega_{N}\left(E\right)$ is the volume of the energy shell
in the phase space $\Omega_{N}\left(E\right)\equiv\int\prod dx_{i}dp_{i}\delta\left(H_{N}\left(\left\{ x_{i},p_{i}\right\} \right)-E\right)$.
We consider here the energy as the only parameter, however generalization
of the following discussion to other quantities conserved by the dynamics
is straightforward.

The only hypothesis of equilibrium statistical mechanics is that averages
with respect to $\mu_{N}$ will correctly describe the macroscopic
behavior of our system. This hypothesis is usually verified after
a sufficiently long time, when the systems has {}``relaxed'' to
equilibrium.

The Boltzmann entropy per particle is defined as \[
S_{N}\left(E\right)\equiv\frac{1}{N}\log\Omega_{N}\left(E\right)\]
 In the following, we will justify that in the long range thermodynamic
limit, the entropy per particle $S_{N}\left(E\right)$ has a limit:
\[
S_{N}\left(E\right)\tod_{N\to\infty}S\left(E\right)\]

The canonical ensemble is defined similarly, using the canonical measure

\[
\mu_{c,N}=\frac{1}{Z_{N}\left(E\right)}\prod_{i=1}^{N}dx_{i}dp_{i}\exp\left[-\beta H_{N}\left(\left\{ x_{i},p_{i}\right\} \right)\right],\]

with the associated partition function $Z_{N}\left(\beta\right)\equiv\int\prod dr_{i}dp_{i}\exp\left[-\beta H_{N}\left(\left\{ x_{i},p_{i}\right\} \right)\right]$
and free energies $F_{N}\left(\beta\right)\equiv\frac{1}{N}\log Z_{N}\left(\beta\right)$
and $F_{N}\left(\beta\right)\tod_{N\to\infty}F\left(\beta\right)$

\subsection{Large deviation results \label{sec:Large-deviations}}

\subsubsection{Justification of the Boltzmann-Gibbs entropy }

\label{sec:Boltzmann-Gibbs}

Let us consider the particle distribution on $\mu-$space
$f\left(x,p\right)$ ($f\left(x,p\right)dxdp$ is the probability to
observe a particle with position $x$ and momentum $p$). $f$ defines a
macrostate as many microscopic states correspond to a given $f$. As
explained in the previous paragraph, the hypothesis of usual
statistical mechanics is that all microscopic states with a given
energy $E$ are equiprobable. Given this uniformity in phase space, we
address the question: what is the number of microscopic states having
the distribution $f$?

It is a classical combinatorial result to show that the logarithm of
number of microscopic states corresponding to a distribution $f$ is
given by \[ s\left[f\right]=-\int dxdp\, f\log f\] where $s$ is
sometimes called the Boltzmann-Gibbs entropy. It is the Boltzmann
entropy associated to the macrostate $f$, in the sense that it counts
the number of microstates corresponding to $f$. We stress that
\textit{no other functional has this probabilistic meaning}, and that
this property is independent of the Hamiltonian.

Thanks to the long range nature of the interaction, for most
configurations, the energy per particle can be expressed in term of
the distribution function $f$, using \begin{equation}
  h\left(f\right)\simd_{N\to\infty}\int\frac{p^{2}}{2}f(x,p)dx~dp~+~\int
  dx_{1}dp_{1}dx_{2}dp_{2}f(x_{1,}p_{1})f(x_{2,}p_{2})V(x_{1}-x_{2}).\,\label{eq:approx_energy}\end{equation}

This mean field approximation for the energy allows to conclude that
the equilibrium entropy is given by

\begin{equation}
S_{N}\left(E\right)=\log\left(\Omega_{N}\left(E\right)\right)\simd_{N\to\infty}NS(E)\,\,\,\textrm{with}\,\,\, S\left(E\right)=\sup_{f}\left\{ s(f)\,\,|\,\, h(f)=E\right\} \label{variationnel}\end{equation}

In the limit of a large number of particles, the mean field approximation
Eq.~(\ref{eq:approx_energy}) and its consequence the variational
problem (\ref{variationnel}) have been justified rigorously for many
systems with long range interactions. The first result assumes a smooth
potential $V$ and has been proved by \cite{Messer_Spohn_1982_JSP},
see also the works by Hertel and Thirring on the self gravitating
fermions~\cite{Hertel_Thirring_1971_CMP}.

\subsubsection{Large deviations}

We explained why the Boltzmann-Gibbs entropy is the correct one to
describe the probability of a given $f$. Large deviations provide a
useful tool to obtain similar results in a wider context. We refer to the very interesting
contributions of Ellis and coworkers
(\cite{EllisHavenTurkington:2000_Inequivalence,EllisHavenTurkington:2002_Nonlinearity_Stability,Ellis_Touchette_Turkington_2004_PA,Boucher_Ellis_1999_AP,TETPhyA2004}).
We also refer to \cite{Barre_Bouchet_DR:2005_JSP} for a simple
detailed explanation of many large deviations results in the context
of long range interacting systems.

In a first step one describes the system at hand by a macroscopic
variable; this may be a coarse-grained density profile $f$, a density
of charges in plasma physics, a magnetization profile for a magnetic
model. In the following, we will generically call this macroscopic
variable $m$; it may be a scalar, a finite or infinite dimensional
variable.

One then associates a probability to each macrostate $m$. Large deviation
theory comes into play to estimate $\Omega(m)$, the number of microstates
corresponding to the macrostate~$m$: \[
\log\left(\Omega_{N}\left(m\right)\right)\simd_{N\to\infty}Ns\left(m\right)~.\]
 This defines the entropy $s(m)$.

In a second step, one has to express the constraints (energy or other
dynamical invariants) as functions of the macroscopic variable $m$.
In general, it is not possible to express exactly $H$; however, for
long range interacting systems, one can define a suitable approximating
mean field functional $h(m)$, as in~Eq.~(\ref{eq:approx_energy}).

Having now at hand the entropy and energy functionals, one can compute
the microcanonical density of states $\Omega(E)$ (\cite{EllisHavenTurkington:2000_Inequivalence}):
the microcanonical solution is simply given by the variational problem
\begin{equation}
\log\left(\Omega_{N}\left(E\right)\right)\simd_{N\to\infty}NS(E)\,\,\,\textrm{with}\,\,\, S\left(E\right)=\sup_{m}\left\{ s(m)\,\,|\,\, h(m)=E\right\} \label{varproblem2}\end{equation}

In the canonical ensemble, similar considerations lead to the conclusion
that the free energy and the canonical equilibrium are given by the
variational problem \begin{equation}
\log\left(Z_{N}\left(\beta\right)\right)\simd_{N\to\infty}NF(\beta)\,\,\,\textrm{with}\,\,\, F\left(\beta\right)=\inf_{m}\left\{ -s(m)+\beta h(m)\right\} \label{varproblem3}\end{equation}

We insist that this reduction of the microcanonical and canonical
calculations to the variational problems (\ref{varproblem2}) and
(\ref{varproblem3}) is in many cases rigorously justified.

\subsection{Ensemble equivalence and simplification of variational problems \label{sec:varproblems}}

As discussed in the previous section, the microcanonical and canonical
equilibrium states are, most of the times, given by
(\ref{varproblem2}) and (\ref{varproblem3}) respectively. These two
variational problems are dual ones: the canonical one is obtained from
the microcanonical one by relaxing a constraint. In the following
section, we discuss the mathematical links between two such dual
variational problems. We then apply this to characterize ensemble
equivalence, and we use it to prove relations between classes of
variational problems.

\subsubsection{Relations between constrained and relaxed variational problems\label{sec:contrainte}}

It is possible to state some general results about the variational
problems~(\ref{varproblem2}) and~(\ref{varproblem3}), independently
of the precise form of the functions $s$ and $h$:

\begin{enumerate}
\item a minimizer $m_{c}$ of (\ref{varproblem3}) is a minimizer of (\ref{varproblem2}),
with constraint $E=h(m_{c})$.
\item a minimizer $m_{\mu}$ of~(\ref{varproblem2}) is a critical point
  of~(\ref{varproblem3}) for some $\beta$, but \emph{it is not always
    a minimizer}: it is a minimizer of~(\ref{varproblem3}) if and only
  if $S$ coincides with its concave hull at point $E=h(m_{\mu})$.
  Otherwise, it may be a local minimum, or a saddle point
  of~(\ref{varproblem3}).
\end{enumerate}

Such results are extremely classical. More detailed results in this
context may be found in
\cite{EllisHavenTurkington:2000_Inequivalence}.  We also refer to
\cite{Bouchet:2007_condmat} for a concise discussion and proof. The
previous points immediately translate into the language of statistical
mechanics, and provide a full characterization of ensemble
inequivalence:

\begin{itemize}
\item A canonical equilibrium is always a microcanonical equilibrium for
some energy~$E$.
\item A microcanonical equilibrium at energy $E$ is a canonical equilibrium
for some temperature $1/\beta$ if and only if $S$ coincides with
its concave hull at energy $E$. Whenever $S$ coincides with its
concave hull, we will say that the ensembles are equivalent; otherwise
we will say they are not equivalent.
\end{itemize}

\subsubsection{Simpler variational problem for statistical equilibria}

In the previous paragraph, we have explained relations between
  solutions of a constrained variational problem and of the associated
  relaxed one. Using similar results and further theoretical
  considerations, it is possible to obtain much simpler variational
  problems than the natural microcanonical ones, for the equilibria of
  Euler and Vlasov equations \cite{Bouchet:2007_condmat}. We think
  that these new results provide essential simplifications that will
  be useful in many studies, we thus describe them in this section.
  However, from a physical point of view, these mathematical results
  may be viewed as technical, and we advise the non expert reader to
  skip this section at first reading.\\

When studying statistical equilibria of systems with long range
interactions, one has to deal with variational problems with one or
several constraints. In the case of the statistical mechanics of the
Euler (resp. the Vlasov equation), there is actually an infinite
number of Casimir's functional conservation laws, encoded in
the initial distribution $d$ of the vorticity field (resp. the particle
distribution function). This is a huge practical limitation. When
faced with real phenomena, physicists can then either give physical
arguments for a given type of distribution $d$ (modeler approach) or
ask whether there exists some distribution $d$ with equilibria close to
the observed field (inverse problem approach). However, in any case
the complexity remains: the class of equilibria is huge.

In the following of this paragraph, we describe recent mathematical
results which allow to relate the microcanonical equilibria to much
simpler variational problems. From a physical point of view, this
simplification is extremely interesting. We describe these results in
the context of the equilibrium theory for the Euler equation (Robert
Sommeria Miller theory
\cite{Miller:1990_PRL_Meca_Stat,SommeriaRobert:1991_JFM_meca_Stat} or
RSM theory), but the following results may be easily generalized to
other cases like the statistical mechanics of the Vlasov equation.  We
refer to \cite{Bouchet:2007_condmat} for a more detailed discussion.
\\

>From a mathematical point of view, one has to solve a microcanonical
variational problem (MVP): maximizing a mixing entropy
$\mathcal{S}[\rho]=-\int_{\mathcal{D}}d^{2}x\int
d\sigma\,\rho\log\rho,$ with constraints on energy $E$ and vorticity
distribution $d$ \[ S(E_{0},d)=\hspace{-0.3cm}\sup_{\left\{
    \rho|N\left[\rho\right]=1\right\} }\hspace{-0.3cm}\left\{
  \mathcal{S}[\rho]\ |\ E\left[\overline{\omega}\right]=E_{0}\
  ,D\left[\rho\right]=d\ \right\} \,\,\mbox{(MVP).}\]
$\rho\left(\mathbf{x},\sigma\right)$ depends on space $\mathbf{x}$ and
vorticity $\sigma$ variables.

During recent years, authors have proposed alternative approaches,
which led to practical and/or mathematical simplifications in the
study of such equilibria. As a first example, Ellis, Haven and
Turkington \cite{EllisHavenTurkington:2002_Nonlinearity_Stability}
proposed to treat the vorticity distribution canonically (in a
canonical statistical ensemble). From a physical point of view, a
canonical ensemble for the vorticity distribution would mean that the
system is in equilibrium with a bath providing a prior distribution of
vorticity. As such a bath does not exist, the physically relevant
ensemble remains the one based on the dynamics: the microcanonical
one. However, the Ellis-Haven-Turkington approach is extremely
interesting as it provides a drastic mathematical and practical
simplification to the problem of computing equilibrium states. A
second example, largely popularized by Chavanis
\cite{Chavanis_Generalized_Entropy_2003,Chavanis_2005PhyD_Generalizedentropy},
is the maximization of generalized entropies. Both the prior
distribution approach of Ellis, Haven and Turkington or its
generalized thermodynamics interpretation by Chavanis lead to a second
variational problem: the maximization of Casimir's functionals, with
energy constraint (CVP)
\[
C(E_{0},s)=\inf_{\omega}\left\{
  \mathcal{C}_{s}[\omega]=\int_{\mathcal{D}}s(\omega)d^{2}x\ |\
  E\left[\omega\right]=E_{0}\ \right\} \,\,\mbox{(CVP)}\] where
$\mathcal{C}_{s}$ are Casimir's functionals, and $s$ a convex function
(Energy-Casimir functionals are used in classical works on nonlinear
stability of Euler stationary flows
\cite{Arnold_1966,Holm_Marsden_Ratiu_Weinstein_1985PhysRev}, and have
been used to show the nonlinear stability of some of RSM equilibrium
states \cite{SommeriaRobert:1991_JFM_meca_Stat,Michel_Robert_1994JSP}).

Another class of variational problems (SFVP), that involve the stream
function only (and not the vorticity), has been considered in relation
with the RSM theory\[ D\left(G\right)=\inf_{\psi}\left\{
  \int_{\mathcal{D}}d^{2}x\,\left[-\frac{1}{2}\left|\nabla\psi\right|^{2}+G\left(\psi\right)\right]\
\right\} \,\,\mbox{(SFVP)}\] Such (SFVP) functionals have been used to
prove the existence of solutions to the equation describing critical
points of (MVP)~\cite{Michel_Robert_1994JSP}. Interestingly, for the
Quasi-geostrophic model, in the limit of small Rossby deformation
radius, such a SFVP functional is similar to the Van-Der-Walls Cahn
Hilliard model which describes phase coexistence in usual
thermodynamics \cite{Bouchet_Sommeria:2002_JFM,Bouchet_These}.  This
physical analogy has been used to make precise predictions in order to
model Jovian vortices
\cite{Bouchet_Sommeria:2002_JFM,Bouchet_Dumont_2003_condmat}.  (SFVP)
functionals are much more regular than (CVP) functionals and thus also
very interesting for mathematical purposes.

When we prescribe appropriate relations between the distribution
function $d$, the functions $s$ and $G$, the three previous
variational problems have the same critical points. This has been one
of the motivations for their use in previous works. However, a clear
description of the relations between the stability of these critical
points is still missing (Is a (CVP) minimizer an RSM equilibria? Or
does an RSM equilibria minimize (CVP)?). This has led to fuzzy
discussions in recent papers. Providing an answer is a very important
theoretical issue because, as explained previously, it leads to deep
mathematical simplifications and will provide useful physical
analogies.

In \cite{Bouchet:2007_condmat} we establish the relation between these
three variational problems. The result is that any minimizer (global
or local) of (SFVP) minimizes (CVP) and that any minimizer of (CVP) is
an RSM equilibria. The opposite statements are wrong in general. For
instance (CVP) minimizers may not minimize (SFVP), but may be instead 
 only saddles. Similarly, RSM equilibria may not minimize (CVP) but be only
saddles, even if no explicit example has yet been exhibited.

These results have several interesting consequences :

\begin{enumerate}
\item As the ensemble of (CVP) minimizers is a sub-ensemble of the ensemble
of RSM equilibria, one can not claim that (CVP) are more relevant
for applications than RSM equilibria.
\item The link between (CVP) and RSM equilibria provides a further
  justification for studying (CVP).
\item Based on statistical mechanics arguments, when looking at the
  Euler evolution at a coarse-grained level, it may be natural to
  expect the RSM entropy to increase. There is however no reason to
  expect such a property to be true for the Casimir's functional. As
  explained above, it may also happen that entropy extrema be (CVP)
  saddles.
\end{enumerate}

\subsection{Classification of phase transitions }

\label{sec:Classification}

Beyond the full characterization of ensemble inequivalence we have
described above, there are many other qualitative features of the
thermodynamics that depend only on the \emph{structure} of the
variational problems (\ref{varproblem2}) and (\ref{varproblem3}).
Indeed, although the precise form of the solution obviously depends on
the problem at hand through the functions $s(m)$ and $h(m)$, it is
possible to \emph{classify} all the different phenomenologies
that one may find in the study of a particular long range interacting
system. The questions in that respect are, increasing complexity at
each step:

\begin{itemize}
\item what are the different possible types of generic points on an entropy
curve $S(E)$ (these correspond to different phases)?
\item what are the possible singular points of a generic $S(E)$ curve (these
correspond to phase transitions)?
\item what are the possible singular points on the $S(E)$ curve, when an
external parameter is varied in addition to the energy (that is how
phase transitions evolve when a parameter is varied)?
\end{itemize}
We address these different levels in the following paragraphs, using
results
from~\cite{EllisHavenTurkington:2000_Inequivalence,Bouchet_Barre:2005_JSP}.
These results are obtained by adapting to the dual variational
problems (\ref{varproblem2}) and (\ref{varproblem3}) ideas that lead
to the Landau classification of phase transitions. In the long range
case however, there is no approximation involved, so the
classification does not suffer from the problems of standard Landau
theory (wrong critical exponents for instance).

\subsubsection{Generic points of an entropy curve}

\begin{figure}
\includegraphics[height=0.2\textheight,keepaspectratio]{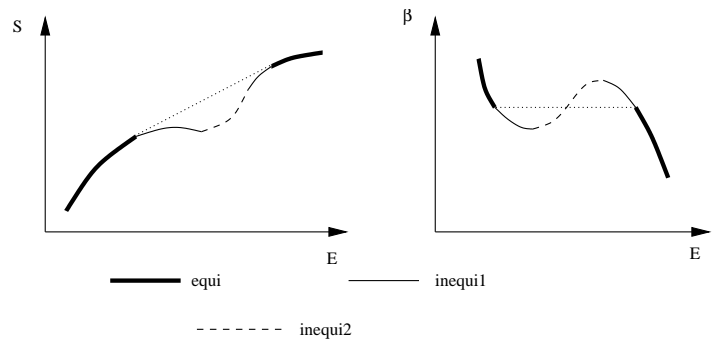}

\caption{The three types of generic points; on the left: entropy $S(E)$ curve,
on the right: caloric $\beta(E)$ curve. Thick, thin and dashed lines
correspond respectively to the three types of points. The dotted lines
shows the Maxwell construction giving the canonical solution in the
inequivalence range.}

\label{fig:generic}
\end{figure}

There are three types of generic points on the entropy curve, see
Fig.~\ref{fig:generic}:

\begin{itemize}
\item Concave points (that is $C_{v}>0$) where canonical and microcanonical
ensembles are equivalent.
\item Concave points where ensembles are not equivalent.
\item Convex points ($C_{v}<0$), where ensembles are always inequivalent.
\end{itemize}

\subsubsection{Singular points of a generic entropy curve: phase transitions}

\begin{figure}
\includegraphics[height=0.2\textheight,keepaspectratio]{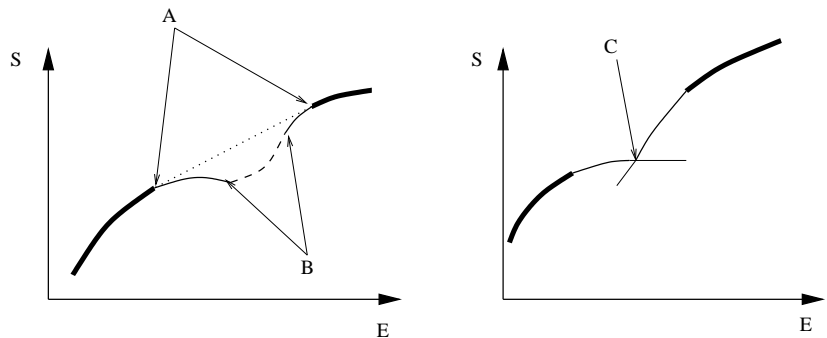}

\caption{The three types of phase transitions (Codimension $0$ singularities),
for a system with no symmetry. \textbf{A:} canonical 1st order transition.
\textbf{B:} canonical destabilization (a local minimum of (\ref{varproblem3})
becomes a saddle point). \textbf{C:} microcanonical 1st order, temperature
jump.}

\label{fig:singular1}
\end{figure}

Generic points as described above define segments of entropy curves,
separated by singular points, that can be of several types. These
points for systems without symmetry are classified in Fig.~\ref{fig:singular1}.

\subsubsection{Singular points on a singular entropy curve }

\begin{figure}
\includegraphics[height=0.3\textheight,keepaspectratio]{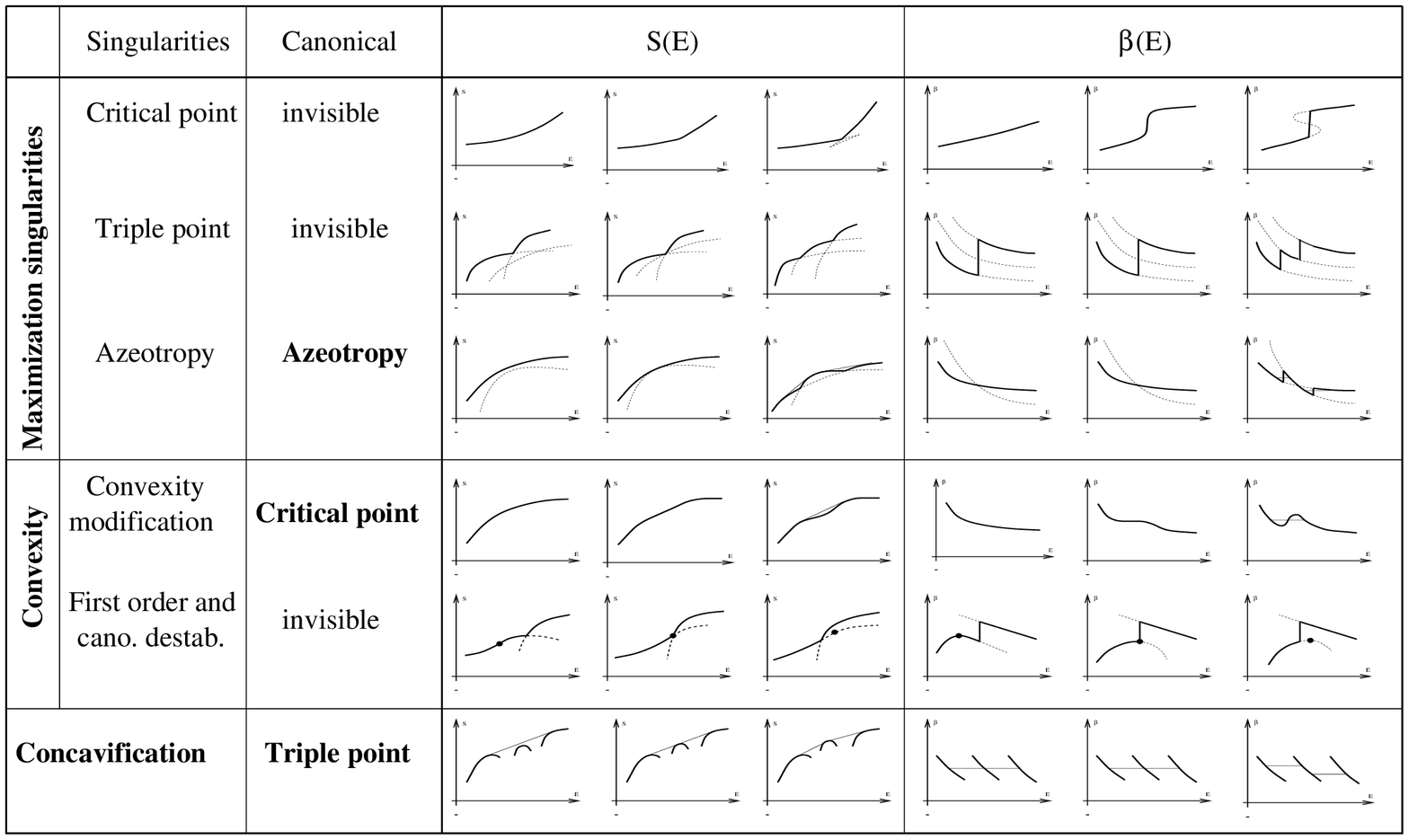}

\caption{Classification of singularities, when one external parameter is varied
(codimension 1 singularities). The first two columns give the singularity
origin and name; the third one gives its status in the canonical ensemble:
invisible means it has no consequence on the canonical solution; the
six following plots are entropic and caloric curves showing the crossing
of the singularity when the external parameter is varied. One recovers
the usual phase transitions (triple point, azeotropy, critical point)
in both ensembles. What is new is the list of the possible links between
the behaviors in each ensemble, and the associated appearance of
ensemble inequivalence. Please see \cite{Bouchet_Barre:2005_JSP}
for a more detailed explanation, and singularities associated with
symmetry breaking.}

\label{fig:singular2}
\end{figure}

When an external parameter is varied, the entropy curve is modified.
Some special values of the parameter correspond to qualitative changes
for the phase transitions. All these possible qualitative changes
are classified in \cite{Bouchet_Barre:2005_JSP}; Fig.~\ref{fig:singular2}
summarizes the results.

All types of phase transitions and ensemble inequivalences found in
the literature so far are reproduced in the classification. In addition,
the classification predicts the possibility of new phenomenologies,
and new routes to ensemble inequivalence, that have not so far been
observed in any specific model.

\subsection{Examples of new phase transitions\label{sub:New_Phases_Transitions}}

In the previous section we explained the existence of many yet
unobserved appearance of ensemble inequivalence, associated to phase
transitions, as described in \cite{Bouchet_Barre:2005_JSP}. More
recently the new finding of two such examples have been
reported~\cite{Venaille_Bouchet_2007_arXiv0710.5606_Fofonoff}:
bicritical points (a bifurcation from a first order phase transition
towards two second order phase transitions) and second order azeotropy
(the simultaneous appearance of two second order phase transitions at
a bifurcation). %
{}We present here these new phase transitions; they are associated
with equilibrium states of the two dimensional Euler equation, when
there is a linear relation between vorticity and stream function. This
was first studied by Chavanis and Sommeria
\cite{ChavanisSommeria:1996_JFM_Classification} in the context of the
Robert-Sommeria-Miller (RSM) statistical mechanics of 2D flows
\cite{Miller:1990_PRL_Meca_Stat,SommeriaRobert:1991_JFM_meca_Stat}.
They found a criterion for the existence of a transition from a
monopole to a dipole above a critical energy, for all (closed) domain
geometry.  In this section, we present an alternative method providing
the same criterion, which generalizes to a large class of
  models, and thus shows the universality of the phenomenon. More
  interestingly this new method clarifies the nature of the phase
transitions involved in this problem and makes the link with the
existence of an ensemble inequivalence
region. 
Those results are presented in a more general context in
  \cite{Venaille_Bouchet_2007_arXiv0710.5606_Fofonoff}, where we note
  the interest of these phase transitions for very simple ocean
  models.

\paragraph{Euler equation and associated variational problem}

Let us consider the 2D Euler equation in a closed domain
$\mathcal{D}$. It can be written as a transport equation for the
vorticity $\omega=\Delta\psi$:
$\partial_{t}\omega+\mathbf{u}.\mathbf{\nabla}\omega=0$. The velocity
field $\mathbf{u}$ is related to $\omega$ via the stream function
$\psi$: $\mathbf{u}=\mathbf{e}_{z}\times\mathbf{\nabla}\psi$, with
$\psi=0$ on $\partial\mathcal{D}$. We introduce the projections
$\omega_{i}$ of the vorticity $\omega$ on a complete orthonormal basis
of eigenfunctions $e_{i}(x,y)$ of the Laplacian: $\Delta
e_{i}=\lambda_{i}e_{i}$, where the $\lambda_{i}$ (all negative) are in
decreasing order.  The stationary states of this equation are
prescribed by a functional relation $\omega=f(\psi)$. In the following
we consider the solutions of the variational problem:
\begin{equation} S(E,\Gamma)=\max_{\omega}\left\{ \mathcal{S}[\omega]\
    |\ \mathcal{E}[\omega]=E\ \&\ \mathcal{C}[\omega]=\Gamma\right\}
  \label{eq:ProblemeVariationnel}\end{equation} 
The
variational problem (\ref{eq:ProblemeVariationnel}) is similar to the
generic problem (\ref{varproblem2}) studied above, with two
constraints instead of one.

\begin{itemize}
\item $\mathcal{S}$ is the entropy of the vorticity field $\omega$; we
  restrict ourselves to a quadratic functional: $\mathcal{S}[\omega]\
  =-\frac{1}{2}\left\langle \omega^{2}\right\rangle
  _{\mathcal{D}}={-\frac{1}{2}\sum\omega}_{i}^{2}$.
\item $\mathcal{E}$ is the total energy:
  $\mathcal{E}[\omega]=\frac{1}{2}\left\langle
    \left(\nabla\psi\right)^{2}\right\rangle
  {}_{\mathcal{\mathcal{D}}}=-\frac{1}{2}\sum_{i}\lambda_{i}\omega_{i}^{2}$
\item $\mathcal{C}$ is the circulation:
  $\mathcal{C}[\omega]=\left\langle \omega\right\rangle
  _{\mathcal{D}}=\sum_{i}\left\langle e_{i}\right\rangle \omega_{i}$
  where $\left\langle e_{i}\right\rangle
  =\int_{\mathcal{D}}e_{i}(x,y)dxdy$
\end{itemize}

To compute critical points of the variational problem
(\ref{eq:ProblemeVariationnel}), we introduce two Lagrange parameters
$\beta$ and $\gamma$, associated respectively with the energy and the
circulation conservation. Those critical points are stationary
solutions for the initial transport equation with
$f\left(\psi\right)=\beta\psi-\gamma$. The solutions of the
variational problem will thus provide the equilibrium states of the
Euler equation that present a linear relationship between vorticity
and stream function, for a given energy and circulation.

The aim of the following paragraphs is to determine which ones among
the critical points are solutions of (\ref{eq:ProblemeVariationnel}).
It will then be possible to draw a phase diagram in the plane $(\Gamma,E)$
for those equilibrium states.

\paragraph{Dual quadratic variational problems}

The problem (\ref{eq:ProblemeVariationnel}), with two constraints,
will be referred to as the microcanonical problem. As already
explained earlier, it is sufficient to study the easier unconstrained
ensembles, unless there is inequivalence of ensembles.  The strategy
is then the following. Start with the easiest problem:

$J(\beta,\gamma)=\min_{q}\left\{ -\mathcal{S}[\omega]\ +\beta\
  \mathcal{E}[\omega]+\gamma\mathcal{C}\left[\omega\right]\right\} $
(grand canonical). Check if all possible values of $E$ or $\Gamma$
correspond to a grand canonical solution; if yes the problem is
solved, otherwise, we turn to the more constrained problem:
$F(\beta,\Gamma)=\min_{q}\left\{ -\mathcal{S}[q]\ +\beta\
  \mathcal{E}[q]\ |\ \mathcal{C}[q]=\Gamma\right\} $ (canonical).

In principle we could eventually have to solve the microcanonical
problem. However, in this case, we will see that the microcanonical
ensemble is equivalent to the canonical one: the whole range of $E$
and $\Gamma$ will be covered by canonical solutions.

We notice first that $\mathcal{S}$, $\mathcal{E}$ are quadratic
functionals and that $\mathcal{C}$ is a linear functional.

We will thus have to look for the minimum of a quadratic functional
with a linear part. Let us call $Q$ the purely quadratic part and
$L$ the linear part of this functional. Then we have three cases

\begin{enumerate}
\item {\small The smallest eigenvalue of $Q$ is strictly positive. The
minimum exists and is achieved by a unique minimizer.}{\small \par}
\item {\small At least one eigenvalue of $Q$ is strictly negative. There
is no minimum.}{\small \par}

\item {\small The smallest eigenvalue of $Q$ is zero (with eigenfunction
$e_{0}$). If $Le_{0}=0$ (case 3a), the minimum exists and each state
of the neutral direction $\left\{ \alpha e_{0}\right\} $ is a minimizer.
If $Le_{0}\neq0$, (case 3b) then no minimum exists. }{\small \par}
\end{enumerate}

\paragraph{The grand canonical ensemble}

\begin{figure}
\resizebox{12truecm}{!}{\includegraphics[width=20cm,keepaspectratio]{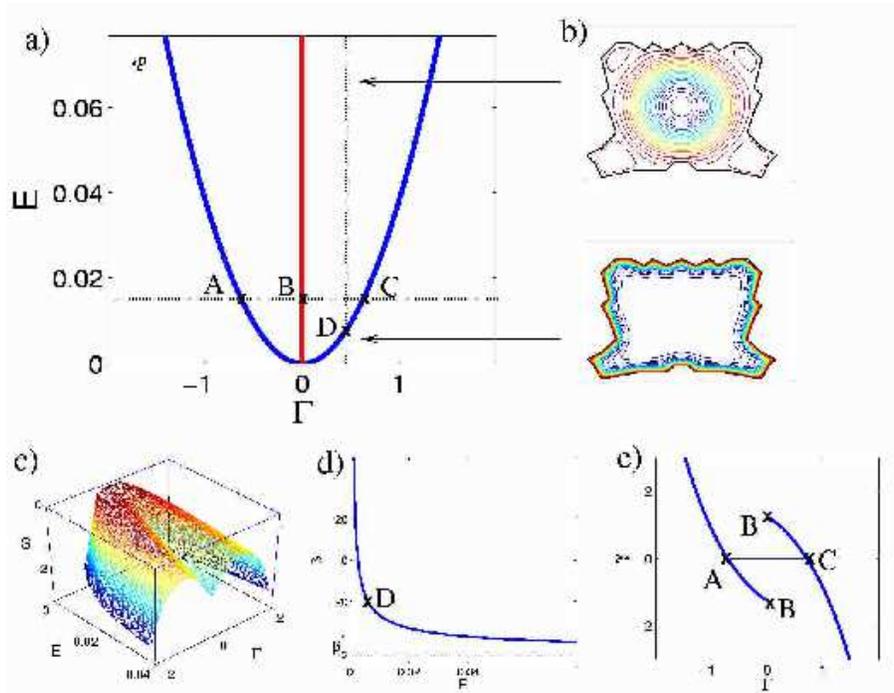}}

\caption{Case i) Phase diagram for a domain having no symmetry, or with a
symmetry axis in the case $\beta^{*}<\beta_{1}^{0}$. a) Internal
parameters are circulation $\Gamma$ and energy $E$. Straight line: first
order transition. Parabola $\mathcal{P}$: above this line there
is no grand canonical solutions; only canonical and microcanonical
solutions exist. It corresponds to an ensemble inequivalence region.
b) examples of flow. c) Equilibrium entropy $S$ as a function of
$E$ and $\Gamma$. We see clearly a region where the concave envelop
of $S(E,\Gamma)$ is not equal to $S(E,\Gamma)$, which includes the
first order transition line; d) $\beta=\partial S/\partial E$ at
fixed $\Gamma$: there is no singularity. e) $\gamma=\partial S/\partial\Gamma$
at fixed $E$: there is a discontinuity of $\gamma$ at $\Gamma=0$
(first order transition). The Maxwell construction shows that the
ensemble inequivalence region is associated with a first order transition
in the ensemble where the value of the energy is fixed.}

\label{Fig:phase_diag_no_trans}
\end{figure}

\begin{figure}
\resizebox{12truecm}{!}{\includegraphics[width=20cm,keepaspectratio]{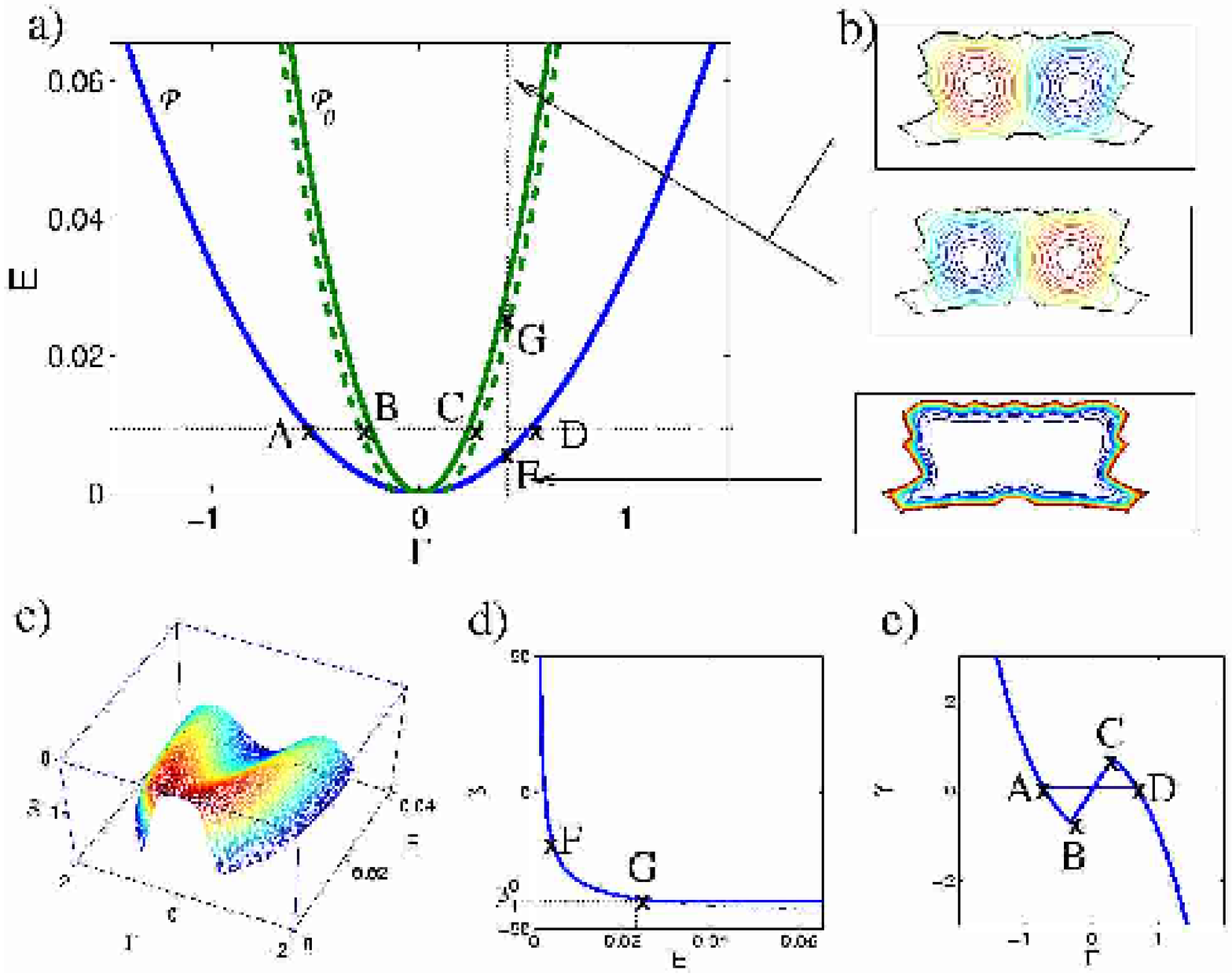}}

\caption{Case ii) Phase diagram for a domain having a symmetry axis, and sufficiently
stretched in a direction perpendicular to this axis (case $\beta^{*}>\beta_{1}^{0}$).
a),b),c),d),e): see figure (\ref{Fig:phase_diag_no_trans}). Dashed parabola $\mathcal{P}_{0}$: second order transition line.
Above this line, $\partial S/\partial E=\beta_{1}^{0}$, and the choice
of one state among two possibilities breaks the system's symmetry.
We see that $\partial\beta/\partial E$ is discontinuous at point
$G$, as well as $\partial\gamma/\partial\Gamma$ at points $B$ and
$C$, which is the signature of a second order phase transition.}

\label{Fig:phase_diag_trans}
\end{figure}

In that case the quadratic operator $Q$ associated to
$\mathcal{J}=-\mathcal{S}\ +\beta\ \mathcal{E}+\gamma\mathcal{C}$ is
diagonal in the Laplacian eigenvector basis. The variational problem
admits a unique solution if and only if $\beta>\lambda_{1}$ (case 1.
above). If $\beta=\lambda_{1}$(case 3. above), a neutral direction
exists if and only if $\gamma=0$. By computing the energy and
circulation of all those states, we prove that there is a unique
solution at each point in the diagram $(E,\Gamma)$, below a parabola
$\mathcal{P}$ (see  figures
\ref{Fig:phase_diag_no_trans} and \ref{Fig:phase_diag_trans}-a).
Because the values of energies above the parabola $\mathcal{P}$ are
not reached, we conclude that there is ensemble inequivalence for
parameters in this region. We then turn to the more constrained
canonical problem to find solutions in this area.

\paragraph{The canonical ensemble}

The circulation is now fixed. We first transform this problem into
an unconstrained variational problem. By using the circulation constraint,
we express one coordinate in term of the others: $\omega_{1}=\left(\Gamma-\sum_{i}\omega_{i}\langle e_{i}\rangle\right)/\left\langle e_{1}\right\rangle $.
This expression is then injected in the functional $\mathcal{F}=-\mathcal{S}+\beta\mathcal{E}$.
The problem is now to find a minimizer $\left\{ \omega_{i}\right\} _{i\ge2}$
of this functional, without constraints. This case requires more computations
that the previous one since the operator $Q$ associated to the quadratic
part of $\mathcal{F}$ is no more diagonal in the basis $\left\{ e_{i}\right\} $.

We first notice that if the domain geometry admits one or more
symmetries, it generically exists eigenfunctions having the property
$\left\langle e_{i}\right\rangle =0$.  In the subspace spanned by all
those eigenfunctions, $Q$ is diagonal, and its smallest eigenvalue is
positive as long as $\beta>\beta_{1}^{0}$, where $\beta_{1}^{0}$ is
the greatest $\lambda_{i}$ on this subspace.  Then we look for the
value of $\beta$ such that the smallest eigenvalue of $Q$ is zero in
the subspace spanned by eigenfunctions with $\left\langle
  e_{i}\right\rangle \neq0$.  Let us call $\beta^{*}$ this value, and
$\omega^{*}$ the corresponding eigenfunction: $Q{[\omega}^{*}]=0$. We
find after some manipulation that $\beta^{*}$ is the greatest zero of
the function $f(x)=1-x\sum_{i\ge1}\langle
e_{i}\rangle^{2}/(x-\lambda_{i})$.  We conclude that there is a single
solution to the variational problem if and only if
$\beta>\max\left(\beta_{1}^{0},\ \beta^{*}\right)$.  When
$\beta=\max\left(\beta_{1}^{0},\ \beta^{*}\right)$, we distinguish two
cases according to the sign of $\beta_{1}^{0}-\beta^{*}$ to discuss
the existence of a neutral direction:

\begin{itemize}
\item {{\small i)}} {\small $\beta_{1}^{0}<\beta^{*}$ we then consider
$\beta=\beta^{*}$. There is a solution (case 3a) if $\Gamma=0$ and
no solution (case 3b) for $\Gamma\neq0$.}{\small \par}
\item {{\small ii)}} {\small $\beta_{1}^{0}>\beta^{*}$ we then consider
$\beta=\beta_{1}^{0}$. There is a solution (case 3a) for all
values of~$\Gamma$.}{\small \par}
\end{itemize}
We thus obtain a criterion, namely the sign of
$\beta_{1}^{0}-\beta^{*}$, that provides two classes of phase
diagrams. This criterion depends only on the domain geometry. If the
domain admits a symmetry axis, and is sufficiently stretched in a
direction perpendicular to this axis, then $\beta_{1}^{0}>\beta^{*}$.
For an ellipse, this is always the case. However, there are domains
with a symmetry axis for which one can find a critical aspect ratio
$\tau_{c}$ that separates the two cases. This is for instance the case
of rectangular domains. For $\tau>\tau_{c}$ , case ii) is realized
($\beta_{1}^{0}>\beta^{*}$) and for $\tau<\tau_{c}$, case i) is
realized ($\beta_{1}^{0}<\beta^{*}$).  To conclude, we find two
classes of phase diagrams:

\begin{itemize}
\item i) Domain without symmetry, or with symmetry with aspect ratio
  $\tau<\tau_{c}$ (see figure \ref{Fig:phase_diag_no_trans}). If
  $\Gamma\ne0$, all energy values are reached in the canonical
  ensemble, and there is a single solution for each point
  $(\Gamma,E)$. If $\Gamma=0$ (straight line on figure
  \ref{Fig:phase_diag_no_trans}), two states coexist for each energy
  value $E$, and $\beta=\beta^{*}$.
\item ii) Domain with symmetry and with an aspect ratio
  $\tau>\tau_{c}$ (see figure \ref{Fig:phase_diag_trans}). There is a
  unique canonical solution at each point of the diagram below a
  parabola $\mathcal{P}_{0}$ in the plane $\left(E,\Gamma\right)$ (see
  the dashed line figure \ref{Fig:phase_diag_trans}). Above this
  parabola, $\beta=\beta_{1}^{0}$, and there is two canonical
  solutions at each point of the diagram. They differ only
  by the sign of the contribution of the Laplacian eigenvector
  associated to the eigenvalue $\beta_{1}^{0}$ (it is a dipole). Below
  the Parabola, this eigenmode has no contribution to the solution.
  The choice of one solution among the two possibilities breaks the
  system's symmetry when $\mathcal{P}_{0}$ is crossed. At high energy,
  the contribution of the dipole is dominant.
\end{itemize}
In both cases we find that all circulation and energy values have
been reached by canonical solutions. We conclude that microcanonical
and canonical ensembles are equivalent: all microcanonical solutions
are also canonical solutions.

\paragraph{Description of phase transitions}

The difference between the two classes of flow is the existence of
either first or second order phase transitions, corresponding
respectively to case i) and ii). The main observation is that first
order and second order microcanonical transitions always take place in
the ensemble inequivalence area. In that respect, those transitions
are signatures of ensemble inequivalence for this class of
two-dimensional flows, which could be observed in laboratory
experiments on quasi two dimensional flows.

\begin{figure}
\resizebox{8truecm}{!}{\includegraphics{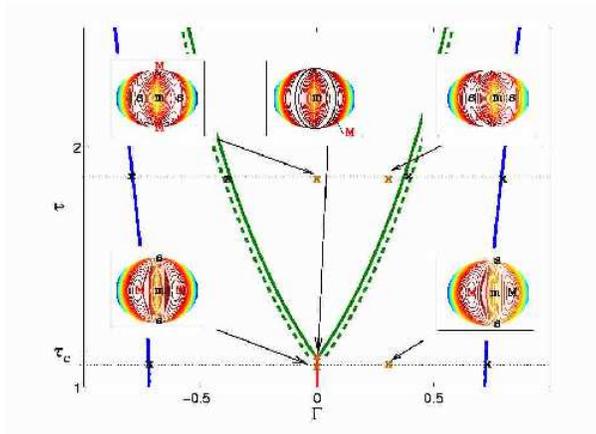}}

\caption{Bicritical point (from Ref.
  \cite{Venaille_Bouchet_2007_arXiv0710.5606_Fofonoff}) for a
  rectangular domain with aspect ratio $\tau$. Bifurcation from a first order transition line to two second order phase transition lines. Insets
  are schematic projections of the Entropy $\mathcal{S}[\omega]$ in a
  plane $(\omega_{1}^{0},\omega^{*})$, by taking into account the
  constraints. M: maximum. m:
  minimum. s: saddle. }

\label{figplotDdep}
\end{figure}

The transition from systems of type i) to systems of type ii), when
the geometry is modified, leads to one of the predicted, but yet
unobserved phase transition of the classification
~\cite{Bouchet_Barre:2005_JSP}, described here in section
\emph{Classification of Phase Transitions}. On figure
\ref{figplotDdep}, we consider a fixed energy, and present the phase
diagram in the $\left(\Gamma,\tau\right)$ plane, where $\tau$ is an
external parameter characterizing the aspect ratio of the domain. In
the microcanonical ensemble, there is a bifurcation from a first order
transition line to two second order transition lines at a critical
value $\tau=\tau_{c}$. Such a bifurcation is referred to as a
bicritical point.

\subsection{Challenges in the equilibrium statistical mechanics\label{sub:Challenges-equilibrium}}

It seems fair to say that the equilibrium statistical mechanics of
long range interacting systems is well understood at a fundamental
level, despite the important differences with the short range case.
However, some challenges and questions remain open; we mention here
some of them.

A first issue is the relevance for natural phenomenon or laboratory
experiments: is it possible to identify situations where equilibrium
statistical mechanics satisfactorily describes the structures observed
(like in Jovian atmosphere~\cite{Bouchet_Sommeria:2002_JFM})?

The main challenge is probably to observe negative specific heat and
ensemble inequivalence in experiments or in real physical systems.  Up
to now, this has not been possible. A first
possibility would be to observe natural phenomena corresponding to
statistical microcanonical equilibria, which are characterized by
negative specific heat, like for instance geophysical flows in
microcanonical situations. Another possibility would be to achieve
this in the lab. One should consider a physical system sufficiently
simple in order to be able to characterize ensemble inequivalence, or
the computation of negative specific heat.  Moreover, exchanges of
energy with the environment have to be negligible over a sufficiently
long time, in order to make sure that the microcanonical ensemble is
the relevant one; this imposes severe constraints on any
laboratory setup. Several candidates have been considered.  One of the
most popular one may be to build synthetic magnetic systems with long
range interactions. Another possibility could be to design simple two
dimensional flow experiments or two dimensional plasma experiments, in
order to reproduce the recently predicted ensemble inequivalence, as
briefly describe
in~\cite{Venaille_Bouchet_2007_arXiv0710.5606_Fofonoff}.

There also several theoretical issues :
\begin{itemize}
\item Could we find physical examples of the new phase transitions found
in the classification?
\item The structure of the dual variational problems
  (\ref{varproblem2}) and (\ref{varproblem3}) appears in other
  physical contexts. Could the results described here, like for
  instance the classification of phase transitions, have an interest,
  when applied to these different situations? For a step in this
  direction, see for instance~\cite{TouchetteBeck2006JSP}.
\item A very interesting and difficult challenge would be to make the
  classification~\cite{Bouchet_Barre:2005_JSP} rigorous.
  This implies to give a precise mathematical definition to the notion
  of a normal form for a mean field variational problem.
\end{itemize}

\section{Kinetic theories of systems with long range interactions \label{sec:Kinetic}}

The previous section provided a brief summary of old and new results
for long range interacting systems at equilibrium. Unfortunately,
it turns out that the coherent structures these systems form, and
the stationary states they reach are generally out of equilibrium.
Although knowledge of equilibrium is a useful benchmark, which usually
yields a qualitative understanding of the physics, some new techniques
are needed to really understand the phenomena at hand. Clearly, we
have to reintroduce the time in our framework and study the dynamics
of the systems.

We have seen that for systems with long range interactions, a mean
field approach is usually exact in the limit of a large number of
particles, when one wants to describe the equilibrium macrostates.
This is valid thanks to an averaging of the potential over many particles.
In the following we explain that a similar mean field approach is
also valid for the dynamics: at each time the potential and the force
can be expressed with a very good approximation from the one particle
distribution function, and thus the BBGKY hierarchy can be safely
truncated.

This well understood fact is the base of the kinetic theory for the
dynamics of systems with long range interaction. This led to the
classical kinetic theories of self gravitating stars, plasmas in the
weak coupling limit, or point vortex models in two dimensional
turbulence. In the limit of a large number of particles, such dynamics
is well approximated by kinetic theories
\cite{Spitzer_1991,Nicholson_1991,Chavanis_houches_2002,Bouchet_Dauxois:2005_PRE,Dubin_ONeil_1988_PhysRevLett_Kinetic_Point_Vortex,Dubin_2003_Phys_Plasma_Collisional_Diffusion_Point_Vortex}:
to leading order in $1/\sqrt{N}$ the dynamics is of a Vlasov type;
after a much longer time, the relaxation towards equilibrium is
governed by Lenard-Balescu type dynamics (or its approximation by the
Landau equation).

In the next subsection, we introduce briefly the Vlasov dynamics, and
the issue of its time of validity. For a large number $N$ of
particles, these systems may exhibit quasi-stationary states (QSS)
\cite{Latora_Rapisarda_Ruffo_1998_PhysRevLett_QSS?,Yamaguchi_Barre_Bouchet_DR:2004_PhysicaA}
(in the plasma or astrophysical context see for instance
\cite{Dubin_ONeil_1999_RevModPhys_Revue_QSS,Spitzer_1991}).  We give a
kinetic interpretation of such states as stable stationary solutions
of the Vlasov
dynamics. 

An interesting question is whether we can predict such
Quasi-Stationary States, from the initial condition of the Vlasov
equation, using statistical mechanics. In the following subsection, we
present recent studies on the equilibrium statistical mechanics of the
Vlasov equation (and not of the N particle dynamics)
\cite{Barre_Dauxois_etal_2004_PRE_FEL,AntoniazziEFR:2006_EPJB_Meca_Stat_Vlasov,Chavanis_PhysicaA_2006},
in the spirit of Lynden-Bell's work \cite{LyndenBell:1968_MNRAS} in
the context of self-gravitating stars.

We then turn to the kinetic theory of these systems beyond the time of
validity of Vlasov equation, which leads to the Lenard-Balescu
equation. This allows to address the important question of the time
scale for the relaxation to equilibrium: this time scale may be of
order $N/\log N$ (this is a classical result by Chandrasekhar for
relaxation to equilibrium of self-gravitating stars, or of a plasma),
of order $N$ (for a smooth potential), or much larger than $N$ (this
is related to the recent result that the Lenard-Balescu operator
identically vanishes for one dimensional
systems~\cite{Bouchet_Dauxois:2005_PRE}).  This last result explains the striking
numerical observation of an $N^{1.7}$ time scale in the HMF
model~\cite{Yamaguchi_Barre_Bouchet_DR:2004_PhysicaA}.

In the following subsection, we explain how a classical kinetic
approach allows to describe the stochastic process of a single
particle in a bath composed by a large number of other particles. This
stochastic process is governed by a usual Fokker-Planck equation. In
classical papers, this bath is at equilibrium. We stress here that
this bath can also be a bath of particles in an out of equilibrium
Quasi Stationary State. We explain recent new results
\cite{Bouchet_Dauxois:2005_PRE} proving that this Fokker Planck
equation has no spectral gap, and lead to long time algebraic
correlations and anomalous diffusion.  This provides a quantitative
prediction for the algebraic autocorrelation function and anomalous
diffusion indices, previously observed in some numerical computations
\cite{Latora_Rapisarda_Ruffo_1999_PhRvL_Superdiffusion,Yamaguchi_2003_PRE,Pluchino_Latora_Rapisarda_2004_PhyD}.
These theoretical predictions have been numerically checked in
\cite{Yamaguchi_Bouchet_Dauxois_2007_JSMTE_Anomalous_Diffusion}.  Some
more recent related results have also been reported in
\cite{Chavanis_Lemou_2005_PRE}.  We note that an alternative
explanation, both for the existence of QSS and for anomalous diffusion
has been proposed in the context of Tsallis non extensive statistical
mechanics
\cite{Latora_Rapisarda_Tsallis_2001_PhRvE,Pluchino_Latora_Rapisarda_2004_PhyD}
(see \cite{Yamaguchi_Barre_Bouchet_DR:2004_PhysicaA} and
\cite{Bouchet_Dauxois:2005_PRE,Bouchet_Dauxois_DR:2006_EurPhysNews}
for further discussions).


The last subsection is devoted to describe some remaining issues and
challenges in the context of the classical kinetic theory for systems
with long range interactions. We also note that we do not describe
many other existing dynamical properties which are common to systems
with long range interactions: vanishing Lyapounov exponents
\cite{Latora_Rapisarda_Ruffo_1998_PhysRevLett_QSS?,Firpo_1998_PhysRevE_Lyapounov},
breaking of ergodicity
\cite{Borgonovi_etal_2004_JStat_Phys_ergodicity,Celardo_Barre_etal_2006_PhRvE_NonconnectivityThreeshold,Bouchet_Dauxois_Mukamel_Ruffo_2007_condmat},
ans so on. All of the common dynamical properties of systems with long
range interactions are a result of similar collective
(self-consistent) dynamics \cite{Del_Castillo_2000_PhyA}.

\subsection{Vlasov dynamics and Quasi-Stationary states\label{sub:Vlasov_QSS}}

As for the equilibrium statistical mechanics, one needs to choose a
scaling to study the kinetic theory; again, the scaling described in
the equilibrium context, which ensures that each particle experiences
a force of order~1, is the appropriate one. The goal is now to
approximate the dynamics of $N$~ordinary differential equations for
the discrete particles dynamics by a single Partial differential
equation for the one-particle distribution function.

For definiteness, we consider the following Hamiltonian system: \begin{equation}
\left\{ \begin{array}{lll}
\dot{x}_{i} & = & p_{i}\\
\dot{p}_{i} & = & -\frac{1}{N}\sum_{j\neq i}\frac{dV}{dx}(x_{i}-x_{j})\end{array}\right.\label{eq:edo}\end{equation}
 The range of the potential $V$ is supposed to be of the same order
of magnitude as the total size of the system: this is our definition
for ''long range interaction''%
\footnote{Albeit rather general, equations~(\ref{eq:edo}) do not include the
2D flows, nor the wave-particles models; most of the following discussion
does apply to these cases too, with small modifications.%
}.

Consider the following continuous approximation of the potential:
\begin{equation}
\Phi(x,t)=\int V(y-x)f(y,p,t)~dy~dp~;\label{eq:phi}\end{equation}
 and the corresponding equation for the one-particle distribution
function $f$ (this is the Vlasov equation):

\begin{equation}
\frac{\partial f}{\partial t}+p\frac{\partial f}{\partial x}-\frac{\partial\Phi}{\partial x}\frac{\partial f}{\partial p}=0\label{eq:vlasov}\end{equation}

Replacing the true discrete potential by $\Phi$ neglects correlations
between particles and finite-$N$ effects. However, as each particle
interacts at any time with an extensive number of other particles,
one may hope that this mean field approach correctly reproduces the
potential experienced by a particle, and becomes exact in the infinite
$N$ limit. Under some regularity assumptions for the potential $V$,
this is indeed correct, and it has been rigorously proved (see \cite{Braun_Hepp_CommMathPhys_1977}
for a very regular $V$, \cite{Hauray_Jabin_ARMA_2007} for a mildly
singular potential). To be more precise, these theorems state the
following: take a discrete $N$-particles initial condition and an
initial continuous one-particle distribution function $f(x,p,0)$
which is close, in some sense, to the former; evolve the $N$ particles
according to~(\ref{eq:edo}), and evolve $f(x,p,0)$ according to~(\ref{eq:phi})
and~(\ref{eq:vlasov}); then the $N$-particles dynamics and $f(x,p,t)$
will remain close for a time at least of the order of $\log N$. Several
remarks are in order:

\begin{enumerate}
\item This implies that if the $t\to\infty$ limit is taken for a fixed
$N$, finite-$N$ effects will come into play; the evolution will
then depart from the Vlasov equation, and we expect the system to
eventually reach the statistical equilibrium. However, for any finite
time $T$ there exists some $N$ such that the system approximately
follows the Vlasov equation up to time $T$.
\item The $\log N$ is optimal in the sense that there exist initial conditions
such that the discrete~(\ref{eq:edo}) and Vlasov~(\ref{eq:vlasov})
dynamics diverge on such a time scale (see \cite{Jain_Bouchet_Mukamel_2007_condmat}
for further discussion). %
\footnote{A recent consideration of the thermodynamic stability of a mean field
Ising model with stochastic dynamics has found the relaxation time
to be logarithmic in $N$ \cite{Mukamel-Ruffo_Schreiber_2005PhRvL}.%
}
\item However, this ''coincidence time'' may in some cases be much longer:
for instance, discrete initial conditions close to a stable stationary
state of the Vlasov equation stay so for a time algebraic in~$N$
(see \cite{Yamaguchi_Barre_Bouchet_DR:2004_PhysicaA} for a numerical
observation and~\cite{Caglioti_Rousset_2007_JStatPhys_QSS} for
a mathematical investigation of the phenomenon).
\item The analogous result for 2D flows is the convergence of the dynamics
of discrete vortices to the corresponding continuous partial differential
equation for the vorticity field (Euler, Quasi-geostrophic...). For
2D flows however, the fundamental equation is the PDE, contrary to
the classical particles case. A mathematical proof of convergence
is given in~\cite{GoodmanHouLowengrub}. For wave-particles systems,
the analogous theorem is given in~\cite{FirpoElskens}.
\item The mathematical proofs cited above do not include the gravitational
and electrostatic $1/r$ cases. It seems however reasonable to believe
that some convergence result towards the Vlasov equation still holds
in this case; Vlasov equation is routinely used by physicists for
these potentials.
\end{enumerate}
In the light of the previous remarks, and if the number of particles
$N$ is big enough, the following dynamical scenario now seems reasonable:

\begin{itemize}
\item Starting from some initial condition, the $N$-particles system approximately
follows the Vlasov dynamics, and evolves on a time scale of order~1.
\item It then approaches a stable stationary state of the Vlasov equation;
the Vlasov evolution stops.
\item Because of discreteness effects, the system evolves on a time scale
of order $N^{\alpha}$ for some $\alpha$, and slowly approaches the
full statistical equilibrium, moving along a series of stable stationary
states of the Vlasov equation.
\end{itemize}
In this scenario, the $N$-particles system gets trapped for long
times out of equilibrium, close to stable stationary states of the
Vlasov equation: these are then called {}``quasi stationary states''
in the literature. This is the basis for the {}``violent relaxation''
theory of Lynden-Bell~\cite{LyndenBell:1968_MNRAS}; Refs.~\cite{Barre_Dauxois_etal_2004_PRE_FEL,Yamaguchi_Barre_Bouchet_DR:2004_PhysicaA,Taruya_Sakagami_PRL_2003}
give examples of this scenario for wave-particles, the HMF and astrophysical
models. The next problem is then to study the stable stationary states
of the Vlasov equation, that is the candidates for the {}``quasi
stationary states''. Before turning to this in the next paragraph,
let us note that there is however no reason for this scenario to be
the only possibility: for instance, the Vlasov dynamics may converge
to a periodic solution of the Vlasov equation~\cite{Firpo_etal_PRE_2001}.

The Vlasov equation (as well as the Euler equation and its variants)
has many invariants: beside the energy $H[f]$ (and possibly the linear
or angular momentum), inherited from the discrete Hamiltonian equations,
the following quantities $C_{s}[f]$, sometimes called Casimirs, are
conserved \emph{for any function} $s$: \begin{equation}
C_{s}[f]=\int s\left(f(x,p,t)\right)~dx~dp~.\label{eq:casimirs}\end{equation}
 Using these invariants, it is possible to construct many stationary
states of the Vlasov equation. Consider the following variational
problem, for a concave function~$s$: \begin{equation}
\sup_{f}\left\{ \int s\left(f(x,p)\right)~dx~dp~~|~\int f~dx~dp=1~~,~~H[f]=e\right\} ~.\label{eq:varproblem}\end{equation}
 Any solution of this variational problem yields a stationary solution
of the Vlasov equation. In addition, the variational structure of
the construction is very useful to study the stability of such states
(see~\cite{EllisHavenTurkington:2002_Nonlinearity_Stability,Holm_etal_PhysRep_1985}
for more details). There is no constraint on the concave function
$s$, so that we have very many stable stationary states of the Vlasov
equation. As a consequence, many numerical or experimental results
can be fitted with a good choice of $s$; this is also a serious limit
of the theory: without a recipe to choose the right stationary state,
the theory is not predictive. We address this problem in the next
paragraph.

We have explained that any Vlasov stable stationary solution is a
Quasi Stationary State. Then, because inhomogeneous Vlasov stationary
states do exist, one should not expect Quasi Stationary States to be
homogeneous. This is illustrated in the case of several
generalizations of the HMF model in
Ref.~\cite{Jain_Bouchet_Mukamel_2007_condmat}.

The issue of the robustness of QSS when the Hamiltonian is perturbed
by short range interactions \cite{Campa_Giansanti_Mukamel_Ruffo_2006_PhyA}
or when the system is coupled to an external bath \cite{BaldovinOrlandini:2006_PRL_ThermalBath}
has also been addressed, and it was found that while the power law
behavior survives, the exponent may not be universal.

\subsection{Lynden Bell statistical mechanics\label{sub:Lynden-Bell}}

Under the Vlasov dynamics, the distribution function $f$ is advected,
by a field which itself depends on $f$. The conservation of Casimirs
amounts to the conservation of the area of all level sets \[
I_{[a,b]}=\{(x,p)~\mbox{such that}~a\leq f(x,p,t)\leq b\}~;\]
 as time goes by the sets are filamented down to a finer and finer
scale, and the filaments get interwoven. Understanding the long time
behavior of this complicated dynamics is not an easy task, analytically
or numerically. Assuming that $f$ tends to one of the many stable
stationary states of the Vlasov equation, the Lynden-Bell statistical
mechanics is a recipe to choose the right one. In a nutshell, at fixed
energy (and possibly linear or angular momentum), it selects the most
mixed state compatible with the Casimirs conservation. It is a maximum
entropy theory; the Lynden-Bell equilibrium is given by the solution
of a problem like~(\ref{eq:varproblem}), the function $s$ being
determined by probability theory and the initial distribution $f(x,p,t=0)$
\footnote{We have to mention that the determination of the Lynden-Bell equilibrium
is in general a difficult task; the calculations are usually practical
only for two- or three-levels initial distributions.%
}.

The idea goes back to a pioneering work of Lynden-Bell in the context
of astrophysics in 1967~\cite{LyndenBell:1968_MNRAS}; the problem
was later revisited by Chavanis and collaborators~\cite{Chavanis_etal_APJ_1996},
in connexion with the statistical mechanics of 2D flows. Let us note
that the analog of Lynden-Bell theory for the Euler and Euler-like
equations of 2D flows is the Robert-Sommeria-Miller theory~\cite{Robert:1991_JSP_Meca_Stat,Miller:1990_PRL_Meca_Stat,SommeriaRobert:1991_JFM_meca_Stat}:
it relies on the very same ideas.

The Lynden-Bell and Robert-Sommeria-Miller theories have had important
successes; let us mention here the descriptions of the core of elliptical
galaxies, and the giant vortices in Jupiter's atmosphere~\cite{Bouchet_Sommeria:2002_JFM}.
However, this is the exception rather than the rule. A lot of works
have been devoted to checking these theories in different contexts,
to which we do not do justice here. To summarize them very briefly,
the rule is that the phase space mixing induced by the Vlasov equation
is not strong enough, so that the theoretical predictions are in general
at best qualitatively correct (see~\cite{Chavanis_PhysicaA_2006}
for a discussion of these issues; see also~\cite{Arad_Johansson_MNRAS_2005}).

\subsection{Order parameter fluctuations and Lenard-Balescu equation}

In this section, we explain briefly how one classically obtains exact
expressions for the $1/\sqrt{N}$ fluctuations of the order parameter,
for a system with long range interactions close to a Quasi Stationary
State. In order to make this discussion as simple as possible, we
treat the case of the HMF model, a one dimensional system with a
smooth two body potential $V$. We follow
\cite{Bouchet_Dauxois:2005_PRE}, and refer to
\cite{Lifshitz_Pitaevskii_1981_Physical_Kinetics} for a plasma physics
treatment, to
\cite{Dubin_ONeil_1988_PhysRevLett_Kinetic_Point_Vortex,Dubin_2003_Phys_Plasma_Collisional_Diffusion_Point_Vortex,Chavanis_houches_2002}
for the case of point vortices and to Ref.
\cite{Binney_Tremaine_1987_Galactic_Dynamics} for self-gravitating
stars.

One could use an asymptotic expansion of the BBGKY hierarchy, where
$1/\sqrt{N}$ is the small parameter, and obtain the same results.
The $1/\sqrt{N}$ fluctuations would then have been obtained by explicitly
solving the dynamical equation for the two point correlation function,
while truncating the BBGKY hierarchy by assuming a Gaussian closure
for the three point correlation function. Such a procedure is justified
in the large $N$ limit (see Ref. \cite{Nicholson_1991}). Our presentation
rather follows the Klimontovich approach.

The state of the $N$-particles system can be described by the {\em
  discrete} single particle time-dependent density function
$f_{d}\left(t,x,p\right)=\frac{1}{N}\sum_{j=1}^{N}\delta\left(x-x_{j}\left(t\right)\right)\delta\left(p-p_{j}\left(t\right)\right),$
where $\delta$ is the Dirac function, $(x,p)$ the Eulerian coordinates
of the phase space and $(x_{i},p_{i})$ the Lagrangian coordinates of
the particles. The dynamics is thus described by the Klimontovich's
equation \cite{Nicholson_1991}.
\begin{equation} \frac{\partial
    f_{d}}{\partial t}+p\frac{\partial f_{d}}{\partial
    x}-\frac{dV}{dx}\frac{\partial f_{d}}{\partial
    p}=0,\label{equationfdiscrete}
\end{equation}
where the potential $V$ that affects all particles is
$V(t,x)\equiv-\int_{0}^{2\pi}\!\!  dy\int_{-\infty}^{+\infty}\!\! dp\
\cos(x-y)\, f_{d}(t,y,p)$. This description of the Hamiltonian
dynamics derived from (\ref{eq:hamiltonian}) is exact : as the
distribution is a sum of Dirac functions it contains the information
on the position and velocity of all the particles.  It is however too
precise for usual physical quantities of interest but will be a key
starting point for the derivation of approximate equations, valid in
the large $N$ limit and describing average quantities.

When $N$ is large, it is natural to approximate the discrete density
$f_{d}$ by a continuous one $f\left(t,x,p\right)$. Considering an
ensemble of microscopic initial conditions close to the same initial
macroscopic state, one defines the statistical average $\langle f_{d}\rangle=f_{0}(x,p)$,
whereas fluctuations of probabilistic properties are of order $1/\sqrt{N}$.
We will assume that $f_{0}$ is any stable stationary solution of
the Vlasov equation. The discrete time-dependent density function
can thus be rewritten as $f_{d}(t,x,p)=f_{0}(x,p)+\delta f(t,x,p)/\sqrt{N}$,
where the fluctuation $\delta f$ is of zero average. We define similarly
the averaged potential $\langle V\rangle$ and its corresponding fluctuations
$\delta V(t,x)$ so that $V(t,x)=\langle V\rangle+\delta V(t,x)/{\sqrt{N}}$.
Inserting both expressions in Klimontovich's equation~(\ref{equationfdiscrete})
and taking the average, one obtains \begin{eqnarray}
\frac{\partial f_{0}}{\partial t}+p\frac{\partial f_{0}}{\partial x}-\frac{d\langle V\rangle}{dx}\frac{\partial f_{0}}{\partial p} & = & \frac{1}{N}\left\langle \frac{d\delta V}{dx}\frac{\partial\delta f}{\partial p}\right\rangle .\label{equationpourfzero}\end{eqnarray}
 The lhs is the Vlasov equation. The exact kinetic equation~(\ref{equationpourfzero})
suggests that the quasi-stationary states of sections \ref{sub:Vlasov_QSS}
and \ref{sub:Lynden-Bell} do not evolve on time scales much smaller
than $N$; this would explain the extremely slow relaxation of the
system towards the statistical equilibrium.


Let us now concentrate on stable homogeneous distributions $f_{0}(p)$,
which are stationary since $\langle V\rangle=0$. Subtracting
Eq.~(\ref{equationpourfzero}) from Eq.~(\ref{equationfdiscrete}) and
using $f_{d}=f_{0}+\delta f/\sqrt{N}$, one gets {\small
  \begin{eqnarray*} \frac{\partial\delta f}{\partial
      t}+p\frac{\partial\delta f}{\partial x} & - & \frac{d\delta
      V}{dx}\frac{\partial f_{0}}{\partial
      p}=\frac{1}{\sqrt{N}}\biggl[\frac{d\delta
      V}{dx}\frac{\partial\delta f}{\partial p}-\left\langle
      \frac{d\delta V}{dx}\frac{\partial\delta f}{\partial
        p}\right\rangle \biggr].\end{eqnarray*} }
For times much shorter than $\sqrt{N}$, we may drop the rhs
encompassing quadratic terms in the fluctuations. The fluctuating part
$\delta f$ are then described, by the linearized Vlasov equation (this
is another result of the Braun and Hepp theorem
~\cite{Braun_Hepp_CommMathPhys_1977,Spohn_1991}).  This suggests to
introduce the spatio-temporal Fourier-Laplace transform of $\delta f$
and $\delta V$. This leads to \begin{eqnarray} \widetilde{\delta
    V}(\omega,k)=-\frac{\pi\left(\delta_{k,1}+\delta_{k,-1}\right)}{\,\varepsilon(\omega,k)}\int_{-\infty}^{+\infty}\!\!
  dp\ \frac{\widetilde{\delta
      f}(0,k,p)}{i(pk-\omega)},\label{fouriertransformofdeltaVsuite}\end{eqnarray}
where \begin{equation}
  \varepsilon(\omega,k)=1+\pi{k}\left(\delta_{k,1}+\delta_{k,-1}\right)\int_{-\infty}^{+\infty}\!\!
  dp\ \frac{{\displaystyle \frac{\partial f_{0}}{\partial
        p}}}{(pk-\omega)}\end{equation} is the dielectric
permittivity. The evolution of the potential autocorrelation, can
therefore be determined. For homogeneous states, by symmetry,
$\langle\widetilde{\delta V}(\omega_{1},k_{1})\widetilde{\delta
  V}(\omega_{2},k_{2})\rangle=0$ except if $k_{1}=-k_{2}=\pm1$.

\subsubsection{Autocorrelation of the potential}

One gets, after a transitory exponential decay, the general result
\begin{equation}
\left\langle {\delta V}(t_{1},\pm1){\delta V}(t_{2},\mp1)\right\rangle =\frac{\pi}{2}\int_{{\cal C}}d\omega\ e^{-i\omega(t_{1}-t_{2})}\ \frac{f_{0}(\omega)}{\left|\varepsilon(\omega,1)\right|^{2}}.\label{correldeltaVfinal}\end{equation}
 This is an exact result, no approximation has been done yet.

\subsubsection{Lenard Balescu equation\label{sub:Lenard-Balescu}}

A similar, but longer, calculation allows to compute the rhs. of Eq.~(\ref{equationpourfzero}),
at order $1/N$. This is very interesting as it gives access to
the slow evolution of the distribution $f_{0}$ due to the {}``collisional''
effects. This is, for systems with long range interactions, the analogue
of the Boltzmann equation for dilute system with short range interactions.
We do not describe the computation in details (see \cite{Lifshitz_Pitaevskii_1981_Physical_Kinetics,Nicholson_1991}),
as we just want to discuss qualitatively the collision operator. This
collision operator is called the Lenard Balescu operator and it leads
to the Lenard Balescu equation.

For system of particle with long range interactions given by a two
body potential $\frac{1}{N}V\left(\mathbf{x}_{1}-\mathbf{x}_{2}\right)$,
the Lenard Balescu equation reads : \begin{equation}
\frac{\partial f_{0}(\mathbf{p},t)}{\partial t}=-\frac{1}{N}\frac{\partial}{\partial\mathbf{p}}.\left[\int d\mathbf{k}d\mathbf{p}'\frac{\phi(k)}{\left|\epsilon(k,\mathbf{k}.\mathbf{p}')\right|}\mathbf{k}.\left(f_{0}\left(\mathbf{p}\right)\frac{\partial f_{0}}{\partial\mathbf{p}}(\mathbf{p}')-f_{0}\left(\mathbf{p}'\right)\frac{\partial f_{0}}{\partial\mathbf{p}}(\mathbf{p})\right)\delta\left(\mathbf{k}.\left(\mathbf{p}-\mathbf{p'}\right)\right)\right],\label{eq:Lenard_Balescu}\end{equation}
 where $\mathbf{k}$ is a wave vector, $\phi(k)$ is the Fourier transform
of the potential $V(\mathbf{x})$, and $\left|\epsilon(k,\mathbf{k}.\mathbf{p}')\right|$
is the dielectric permittivity. One note that this is a quadratic
operator, as for the Boltzmann equation. Moreover, this operator involve
a resonance condition in the Dirac distribution $\delta\left(\mathbf{k}.\left(\mathbf{p}-\mathbf{p'}\right)\right)$.

>From this equation one clearly expects a relaxation towards equilibrium
of any Quasi-Stationary state with a characteristic time of order
$N$. We note that for plasma or self gravitating systems, due to
the small $r$ divergence of the interaction potential, the Lenard
Balescu operator diverges at small scales. This is regularized by
close two body encounters, fixing a small scale cutoff. This leads
to a logarithmic correction to the relaxation time, which is then
the Chandrasekhar time proportional to $\log(N)/N.$

One clearly sees on equation (\ref{eq:Lenard_Balescu}) that the
mechanism for evolution of the distribution function is related to the
resonances of two particles. An essential point is that the condition
$\mathbf{k}.\left(\mathbf{p}-\mathbf{p'}\right)=0$ cannot be fulfilled
for one dimensionnal systems. It would indeed imply $p=p'$, and
because the Lenard Balescu operator is odd in the variable $p$, it
will vanish. Another way to obtain the same result, is to directly
compute the rhs of Eq.~(\ref{equationpourfzero}).  We do not report
such long and tedious computations, but it shows that it identically
vanishes at order $1/N$, for one dimensional systems.

This proves that Vlasov stable distribution function will not evolve
on time scales smaller or equal to $N$. This is an important result:
\textit{generic out of equilibrium distributions, for one dimensionnal
systems, evolve on time scales much larger than $N$}. This is in
agreement with the $N^{1.7}$ scaling law which was numerically reported
\cite{Yamaguchi_Barre_Bouchet_DR:2004_PhysicaA}.

\begin{figure}[htbp]
 \centering \includegraphics[height=0.2\textheight,keepaspectratio]{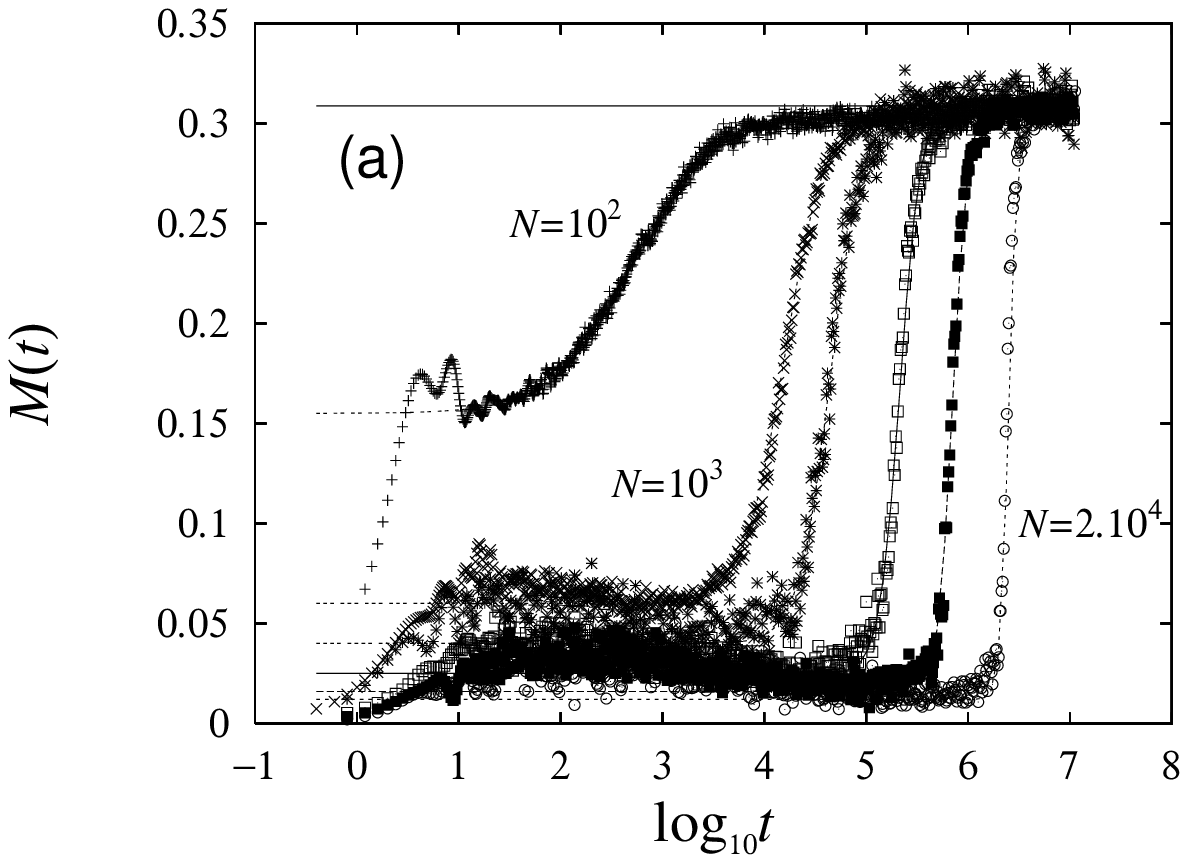}
\includegraphics[height=0.2\textheight,keepaspectratio]{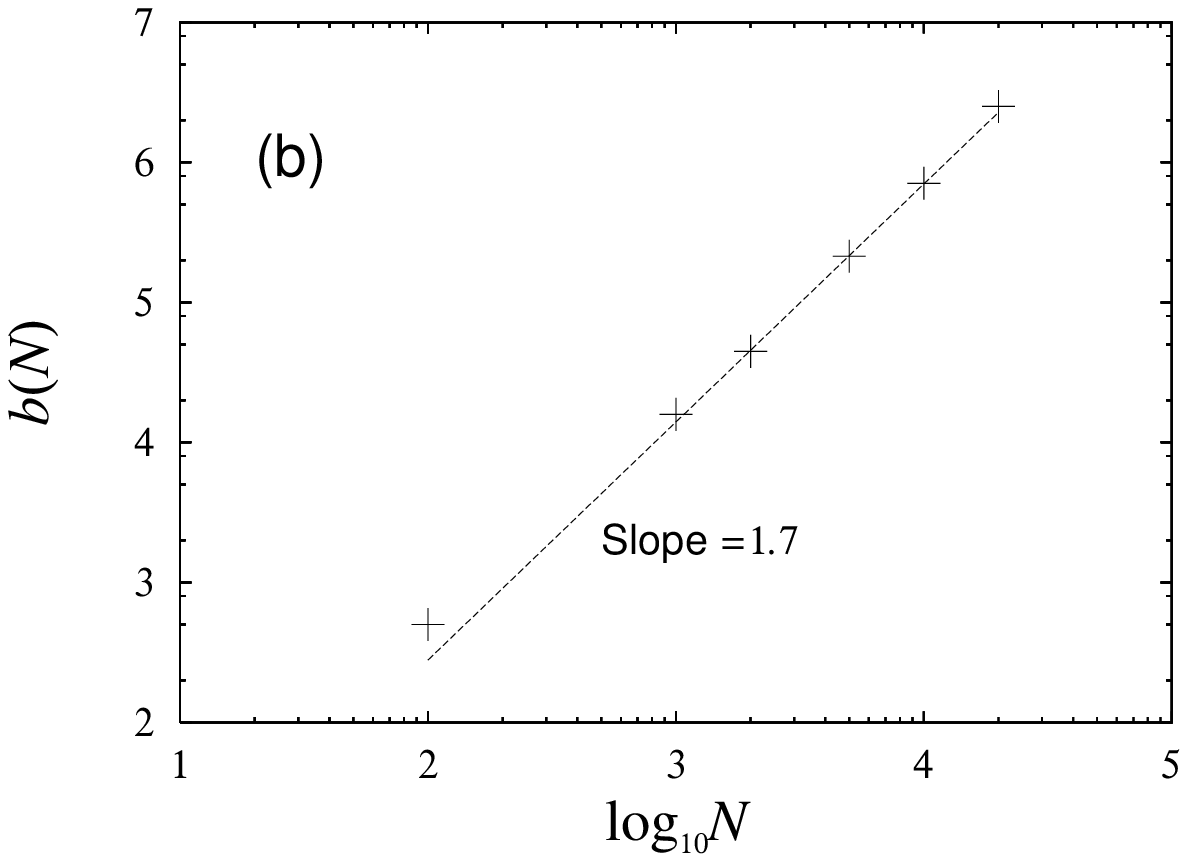}

\caption{Panel (a) presents the temporal evolution of the magnetization~$M(t)$,
for the HMF model, for different particles numbers: $N=10^{2}(10^{3})$,
$10^{3}(10^{2})$, $2.10^{3}(8)$, $5.10^{3}(8)$, $10^{4}(8)$ and
$2.10^{4}(4)$ from left to right, the number between brackets corresponding
to the number of samples. The horizontal line represents the equilibrium
value of $M$. Panel~(b) shows the logarithmic timescale $b(N)$
as a function of $N$, whereas the dashed line represents the law
$10^{b(N)}\sim N^{1.7}$. From Ref. \cite{Yamaguchi_Barre_Bouchet_DR:2004_PhysicaA}.
\label{fig:Magnetization_HMF}}
\end{figure}

\subsubsection{The stochastic process of a single particle in a bath}

Let us now consider relaxation properties of a test-particle, indexed
by 1, surrounded by a background system of $(N-1)$ particles with
a homogeneous distribution. The fluctuation of the potential is thus
\begin{equation}
\delta V(t,x)\equiv-\int_{0}^{2\pi}\!\!\!\!\!\!\! dy\!\!\int_{-\infty}^{+\infty}\!\!\!\!\!\!\!\! dp\
\cos(x-y)\,\delta f(t,y,p)-\frac{1}{\sqrt{N}}\cos\left(x-x_{1}\right).\label{fluctpotentialbis}\end{equation}
 Using the equations of motion of the test particle and omitting the
index $1$ for the sake of simplicity, one obtains $p(t)=p(0)-\int_{0}^{t}\!\! du{(d\delta V(u,x(u)))}/{(dx)}/{\sqrt{N}}$.
By introducing iteratively the expression of $x$ in the rhs and expanding
the derivative of the potential, one gets the result at order $1/N$.
The key point is that this approach does not use the usual ballistic
approximation. As a consequence, we obtain an exact result at order
$1/N$. This is of paramount importance here to treat accurately the
{\em collective effects}. As the changes in the impulsion are small
(of order $1/\sqrt{N}$), the description of the impulsion stochastic
process by a Fokker-Planck equation is valid. This last equation is
then characterized by the time behavior of the first two moments $\langle\left(p(t)-p(0)\right)^{n}\rangle$.
Using the generalization of formula~(\ref{correldeltaVfinal}) when
the effect of the test particle is taken into account, one obtains
in the large $t$-limit \begin{eqnarray}
\langle\left(p(t)-p(0)\right)\rangle & \simd_{t\to+\infty} & \!\!\frac{t}{N}\left(\frac{dD}{dp}(p)+\frac{1}{f_{0}}\frac{\partial f_{0}}{\partial p}D(p)\right)\quad\null\label{differentmoments1}\\
\langle\left(p(t)-p(0)\right)^{2}\rangle & \simd_{t\to+\infty} & \!\!\frac{2t}{N}\, D(p),\label{differentmoments2}\end{eqnarray}
 where the diffusion coefficient $D(p)$ can be written as {\small \begin{equation}
D(p)=2\,\mbox{Re}\int_{0}^{+\infty}\!\!\!\!\!\!\!\! dt\
e^{ipt}\,\left\langle {\delta V}(t,1){\delta V}(0,-1)\right\rangle =\ \frac{{\pi^{2}}f_{0}(p)}{\left|\varepsilon(p,1)\right|^{2}}.\label{diffusioncoefficient}\end{equation}
}  These results are the exact leading order terms in an expansion
where $1/N$ is the small parameter.

Using time variable $\tau=t/N$ as suggested by Eqs.~(\ref{differentmoments1})
and~(\ref{differentmoments2}), the Fokker-Planck equation describing
the time evolution of the distribution of the test particle is \begin{equation}
\frac{\partial f_{1}(\tau,p)}{\partial\tau}=\frac{\partial}{\partial p}\left[D(p)\left(\frac{\partial f_{1}(\tau,p)}{\partial p}-\frac{1}{f_{0}}\frac{\partial f_{0}}{\partial p}f_{1}(\tau,p)\right)\right].\label{fokkerplanckequation}\end{equation}
 We stress that this equation depends on the bath distribution $f_{0}$.
It is valid both for equilibrium and and out of equilibrium $f_{0}$,
provided that $f_{0}$ is a stable stationary solution of the Vlasov
equation. In the limit $\tau\to\infty$ (more precisely $1<<\tau<<N$),
the bracket vanishes: the PDF $f_{1}$ of the test particle converges
toward the quasi-stationary distribution $f_{0}$ of the surrounding
bath. This is in complete agreement with the result that $f_{0}$
is stationary for time scales of order~$N$.

All the results of this section, except the fact that the Lenard
Balescu equation vanishes for one dimensional systems, are classical
results.  In the next section we explain recent results
related to the very interesting and peculiar properties of the
Fokker-Planck equation (\ref{fokkerplanckequation}).

\subsection{Autocorrelation function with algebraic decay and anomalous diffusion\label{sub:Anomalous_diffusion}}

In this subsection, we present recent results \cite{Bouchet_Dauxois:2005_PRE}
which predicted the existence of non exponential relaxation, autocorrelation
of the momentum $p$ with algebraic decay at large time, and anomalous
diffusion of the spatial or angular variable $x$. They clarify the
highly debated disagreement between different numerical simulations
reporting either anomalous \cite{Latora_Rapisarda_Ruffo_1999_PhRvL_Superdiffusion}
or normal \cite{Yamaguchi_2003_PRE} diffusion, in particular by
delimiting the time regime for which such anomalous behavior should
occur. We briefly recall that when the moment of order $n$ of the
distribution scales like $\tau^{n/2}$ at large time, such a transport
is called {\em normal}. However, {\em anomalous} transport \cite{Bouchaud_Georges_1990_PhysRep,Castiglione_Mazzino_Muratore_Vulpiani_1999PhyD_Anomalous_Diffusion},
where moments do not scale as in the diffusive case, were reported
in some stochastic models, in continuous time random walks (Levy walks),
and for systems with a lack of stationarity of the corresponding stochastic
process \cite{Bouchet_Cecconi_Vulpiani:2004_PRL}.\\

These results have been obtained by analyzing theoretically the
properties of the Fokker-Planck equation (\ref{fokkerplanckequation}).
>From the physical point of view, as particles with large momentum $p$
fly very fast in comparison to the typical time scales of the
fluctuations of the potential, they experience a very weak diffusion
and thus maintain their large momentum during a very long time (one
sees from equation (\ref{diffusioncoefficient}), using
$\left|\varepsilon(p,1)\right|^{2}\tod_{p\rightarrow\infty}1$,
that the diffusion coefficient decays as fast as the bath distribution
$f_{0}\left(p\right)$ for large times). Because of this very weak
diffusion for large $p$, the distribution of waiting time for passing
from a large value of $p$ to a typical value of $p$, is a thick
distribution. This explains the algebraic asymptotic for the
correlation function. From a mathematical point of view, these
behaviors are linked to the fact that the Fokker-Planck equation
(\ref{fokkerplanckequation}) has a continuous spectrum down to its
ground state (without gap).  This leads to a non exponential
relaxation of the different quantities and to long-range temporal
correlations \cite{Bouchet_Dauxois:2005_PRE,Bouchet_Dauxois:2005_JOP}.
These results will generalize to the kinetic theory of any system for
which the slow variable (here the momentum) live in an infinite space.

By explicitly deriving an asymptotic expansion of the eigenvalues and
eigenfunctions of the Fokker Planck equation, the exponent for the
algebraic tail of the autocorrelation function of momenta has been
theoretically computed
\cite{Bouchet_Dauxois:2005_PRE,Bouchet_Dauxois:2005_JOP}.  This
mechanism is new in the context of kinetic theory. However, we have
discovered later that similar Fokker-Planck equations, with a rapidly
vanishing diffusion coefficients obtained by other physical
mechanisms, had been studied
\cite{Farago_2000_Euro_Letters_Persistence,Lillo_Micciche_Mantegna_2002cond.mat,Lutz_2004_PhRvL_Ergodicity}.
A more recent alternative approach to the same phenomena has been
proposed \cite{Chavanis_Lemou_2005_PRE}, together with interesting
discussions of kinetic applications.\\

Let us present the results in the context of the HMF model, for which
algebraic large time behaviors for momentum autocorrelations had been
first numerically observed in Refs. \cite{Latora_Rapisarda_Tsallis_2001_PhRvE,Pluchino_Latora_Rapisarda_2004_PhyD}.
In its Quasi Stationary States, the theoretical law for the diffusion
of angles $\sigma_{x}^{2}(\tau)$ has been also derived in \cite{Bouchet_Dauxois:2005_PRE,Bouchet_Dauxois:2005_JOP}.
The predictions for the diffusion properties are listed in Table~\ref{tab:Cp}.

\begin{table}[htbp]
 \centering \begin{tabular}{|l|c|c|c|}
\hline
&
&
&
\tabularnewline
Tails &
$f_{0}(p)$ &
$C_{p}(\tau)$ &
$\sigma_{x}^{2}(\tau)$\tabularnewline
&
&
&
\tabularnewline
\hline
&
&
&
\tabularnewline
Power-law &
$|p|^{-\nu}$ &
$\tau^{-\alpha}$ &
$\tau^{2-\alpha}$ \tabularnewline
&
&
&
\tabularnewline
\hline
&
&
&
\tabularnewline
Stretched exponential &
$\quad\exp(-\beta|p|^{\delta})$\quad{}&
\quad{}$\frac{(\ln\tau)^{2/\delta}}{\tau}$\quad{}&
\quad{}$\tau(\ln\tau)^{2/\delta+1}$\quad{}\tabularnewline
&
&
&
\tabularnewline
\hline
\end{tabular}

\caption{Asymptotic forms of initial distributions $f_{0}(p)$, and theoretical
predictions of correlation functions $C_{p}(\tau)$ and the diffusion
$\sigma_{x}^{2}(\tau)$ in the long-time regime. Asymptotic forms
of the distribution and the predictions are assumed and predicted
in the limits $|p|\to\infty$ and $\tau\to\infty$ respectively, where
$\tau=t/N$ is a rescaled time. The exponent $\alpha$ is given as
$\alpha=(\nu-3)/(\nu+2)$. See Ref. \cite{Bouchet_Dauxois:2005_PRE,Bouchet_Dauxois:2005_JOP}
for details.}

\label{tab:Cp}
\end{table}

When the distribution $f_{0}(p)$ is changed within the HMF model,
a transition between weak anomalous diffusion (normal diffusion with
logarithmic corrections) and strong anomalous diffusion is thus predicted.
We have numerically confirmed the theoretical predictions \cite{Yamaguchi_Bouchet_Dauxois_2007_JSMTE_Anomalous_Diffusion}.
For initial distributions with power-law or Gaussian tails, correlation
function and diffusion are in good agreement with numerical results.
Diffusion is indeed {\em anomalous super-diffusion} in the case
of power-law tails, while {\em normal} when Gaussian. In the latter
case, the system is at equilibrium, but the diffusion exponent shows
a logarithmically slow convergence to unity due to a logarithmic correction
of the correlation function. This long transient time to observe normal
diffusion, even for Gaussian distribution and at equilibrium, suggests
that one should be very careful to decide whether diffusion is anomalous
or not.

\begin{figure}
\includegraphics[height=0.2\textheight,keepaspectratio]{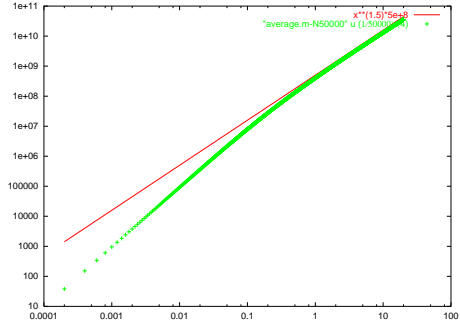}

\caption{Angles diffusion ( $<(x(t)-x(0))^{2}>$ as a function of time)
  in the HMF model, for a quasi-stationary state. Points are from a N
  body numerical simulation, the straight line is the analytic
  prediction by the kinetic theory. For large times
  $<(x(t)-x(0))^{2}>\simd_{t\rightarrow\infty}t^{\nu}$ with
  $\nu\neq1$. Such an unexpected anomalous diffusion is also observed
  at equilibrium (see \cite{Bouchet_Dauxois:2005_PRE} for more
  details)}

\label{fig:diffusion_anomale}
\end{figure}

We note the existence of another interpretation of Quasi Stationary
States and anomalous diffusion : the algebraic behaviors for momentum
autocorrelations have been fitted using q-exponential functions
\cite{Latora_Rapisarda_Tsallis_2001_PhRvE,Pluchino_Latora_Rapisarda_2004_PhyD},
derived from Tsallis' non extensive statistical mechanics. Our
theoretical and numerical results are in disagreement with this
interpretation (see
\cite{Bouchet_Dauxois_DR:2006_EurPhysNews,Bouchet_Dauxois:2005_PRE,Yamaguchi_Bouchet_Dauxois_2007_JSMTE_Anomalous_Diffusion}
for further discussions).  By contrast with the use of non extensive
statistical mechanics, we think that our theory explains the phenomena
of long range temporal correlation and of anomalous diffusion from
first principles.

\subsection{Challenges in kinetic theories\label{sub:Challenges-in-kinetic}}

Our main message in this section is that a classical kinetic theory
approach for these long range interacting systems already explains
many intriguing behaviors of these systems. However, in contrast with
the equilibrium theory, many questions remain open; we mention here
some of them, without any pretention to be exhaustive:

\begin{itemize}
\item Can we find a better recipe than Lynden-Bell's theory to predict the
outcome of the Vlasov evolution? This seems hopeless in a general
setting (see for instance the discussion in~\cite{Chavanis_PhysicaA_2006}).
\item Is it possible to explain the $1.7$ exponent for the relaxation
  to equilibrium in the HMF model, and does it have some universality?
  More generally, is it possible to extract other general features of
  the dynamics beyond the Vlasov equation, like the anomalous
  diffusion, or the long relaxation times described above?
\item At the mathematical level: is it possible to improve on~\cite{Caglioti_Rousset_2007_JStatPhys_QSS}
  concerning the lifetime of QSS? Can the
  convergence theorems to the Vlasov equation be extended to more
  singular potentials?
\item The most important issue concerning kinetic theories is a clear
  understanding of the limits of validity the different equations.
  Whereas, for smaller times, kinetic theory are based on solid
  theoretical arguments, the understanding of larger time behavior of
  an ensemble of trajectories, initially close to one another, is not
  yet understood. Numerical computations could be very useful in order
  to understand that. Very few direct numerical tests of the kinetic
  theories have been performed up to now.  The main reason is probably
  the difficulty for such tests, because of the long time needed for
  such test. We think it would be highly relevant to consider such
  problem, in models as simple as possible.
\end{itemize}

\section{Out of equilibrium \label{sec:NESS}}

\subsection{Motivations}

We have described the computation of equilibrium states for systems
with long range interactions in the first section, and addressed the
problem of relaxation to equilibrium in the second one.  These two
types of problems concern isolated Hamiltonian systems, systems which
may be considered so on the relevant time scales, or systems in
contact with a thermal bath.  In many cases of interest, the system
experiences random forces and dissipation. Very often the mechanism
for dissipation and random forces are from a different origin, and do
not act as a thermal bath.  As a consequence, detailed balance is no
more valid and the system is subject to fluxes of energy or possibly
of other conserved quantities; the average energy of the system is
fixed by the balance between forcing and dissipation. The
understanding of the properties of the corresponding Non Equilibrium
Steady States (NESS) is thus of deep importance. We present here first
studies of such NESS in the context of systems with long range
interactions.  The most prominent result is the finding of out of
equilibrium phase transitions.

These first studies have been done in the context of two dimensional
flows. This is indeed essential in this case, as in many applications
of fluid dynamics, one of the most important problem is the prediction
of the very high Reynolds' large-scale flows. The highly turbulent
nature of such flows, for instance ocean circulation or atmosphere
dynamics, renders a probabilistic description desirable, if not
necessary.  At equilibrium, a statistical mechanics explanation of the
self-organization of geophysical flows has been proposed by
Robert-Sommeria and Miller (RSM). Out of equilibrium, there are
several practical and fundamental problems to understand: How the
invariants are selected by the presence of weak forces and
dissipation? What are the associated fluctuations? Are all forcings
compatible with RSM equilibria?

\textcolor{black}{We will thus study the Navier Stokes equation with
weak random stochastic forces and dissipation: \begin{equation}
\frac{\partial\omega}{\partial t}+{\bf \mathbf{u}}.\nabla\omega=\nu\Delta\omega-\alpha\omega+f_{s}\label{eq:NavierStokes_Stochastic}\end{equation}
 where $\omega$ is the vorticity, $f_{s}$ is a random force, $\alpha\omega$
is the Rayleigh dissipation and $\nu$ is the fluid viscosity.}

\subsection{Out of equilibrium phase transitions}

In many turbulent geophysical flows, one can see transitions, at
random times, between two states with different large scale flows. The
most famous example are probably the time reversal of the earth
magnetic field. We may cite also experimental studies of such
phenomena, for two dimensional magnetic flows
\cite{Sommeria_1986_JFM_2Dinverscascade_MHD}, rotating tank experiment
in relation with weather regimes in meteorology
\cite{Tian_Weeks_etc_Ghil_Swinney_2001_JFM_JetTopography}, or magnetic
field reversal in MHD
\cite{Berhanu_etc_Fauve_2007_EPL_MagneticFieldReversal}. In all these
examples, this generic phenomenon takes place in systems with a large
number of degrees of freedom. The case of simple turbulent flows may
be studied in much details theoretically and numerically; we focus
here on the case of the two dimensional Navier-Stokes equation with a
random force.

\begin{figure}[htbp]
 \centering \includegraphics[height=0.2\textheight,keepaspectratio]{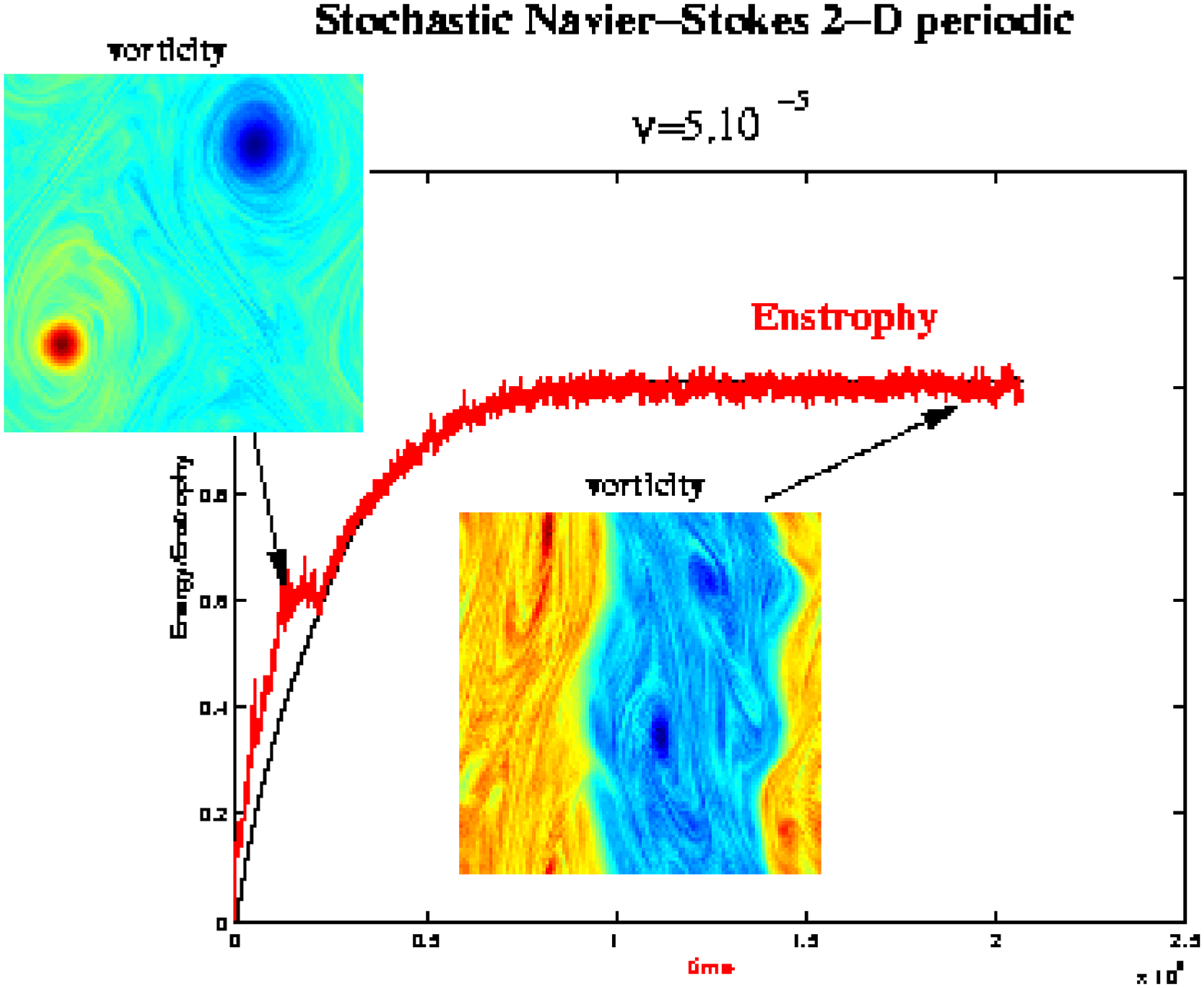}
\includegraphics[height=0.2\textheight,keepaspectratio]{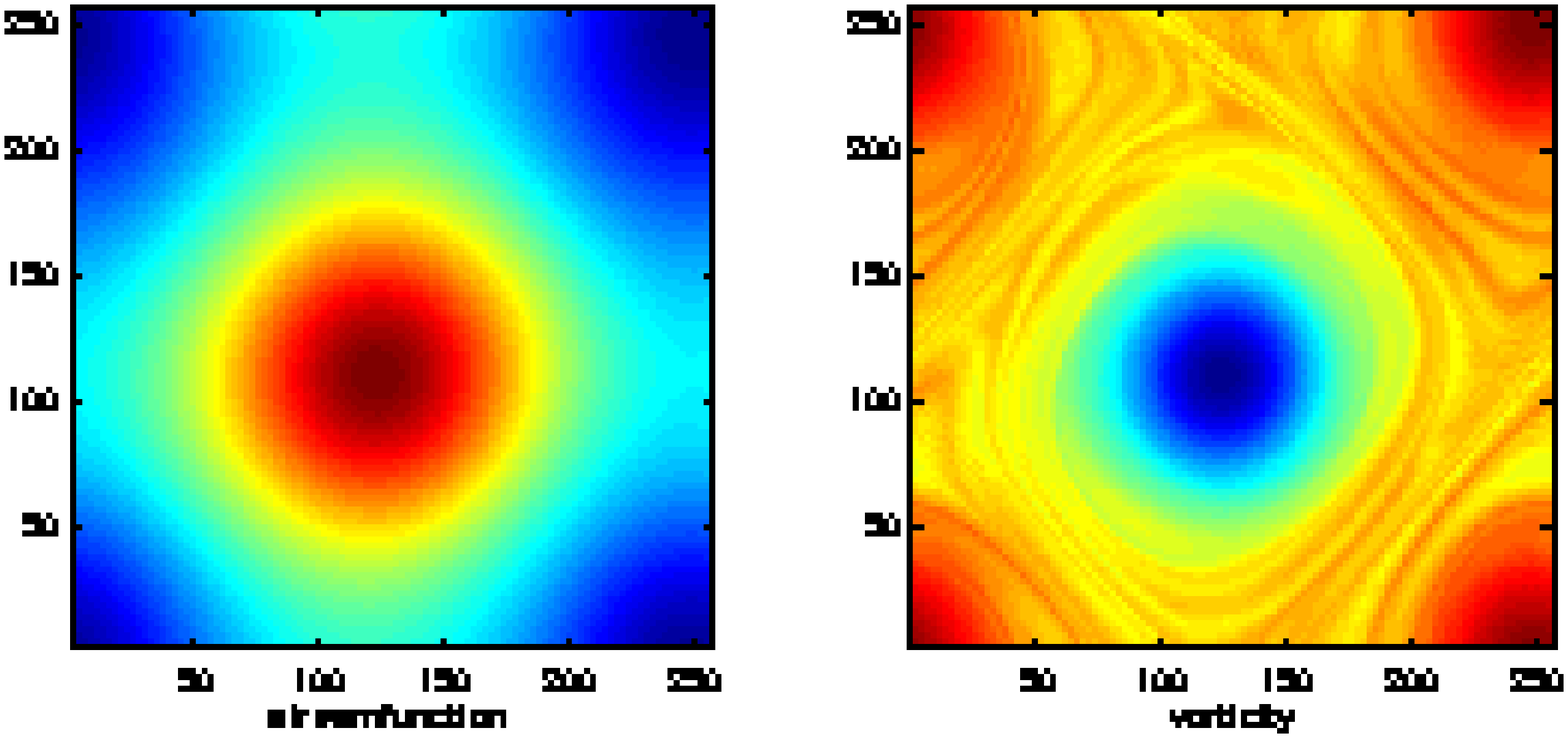}

\caption{Right panel: relaxation towards a statistically stationary state
of the two dimensional Navier Stokes equation. After a transient state
with a dipole vorticity field, the flow switches to a zonal (unidirectional)
organization of the vorticity field. Left panel: for a different value
of the control parameter (here the aspect ratio of the domain), we
observe a dipole organization in a statistically stationary situation.
\label{fig:Phase_Transition_Vorticity}}
\end{figure}

Figure \ref{fig:Phase_Transition_Vorticity} shows the relaxation
of the 2D Navier-Stokes equation to a statistically stationary state.
It illustrates that, depending on the aspect ratio of the domain,
two types of large scale flows are possibly observed, either dipoles
or unidirectional (zonal) flows. We note that these two topologies
are also predicted by the equilibrium statistical theory.

\begin{figure}[htbp]
\centering \includegraphics[height=0.2\textheight,keepaspectratio]{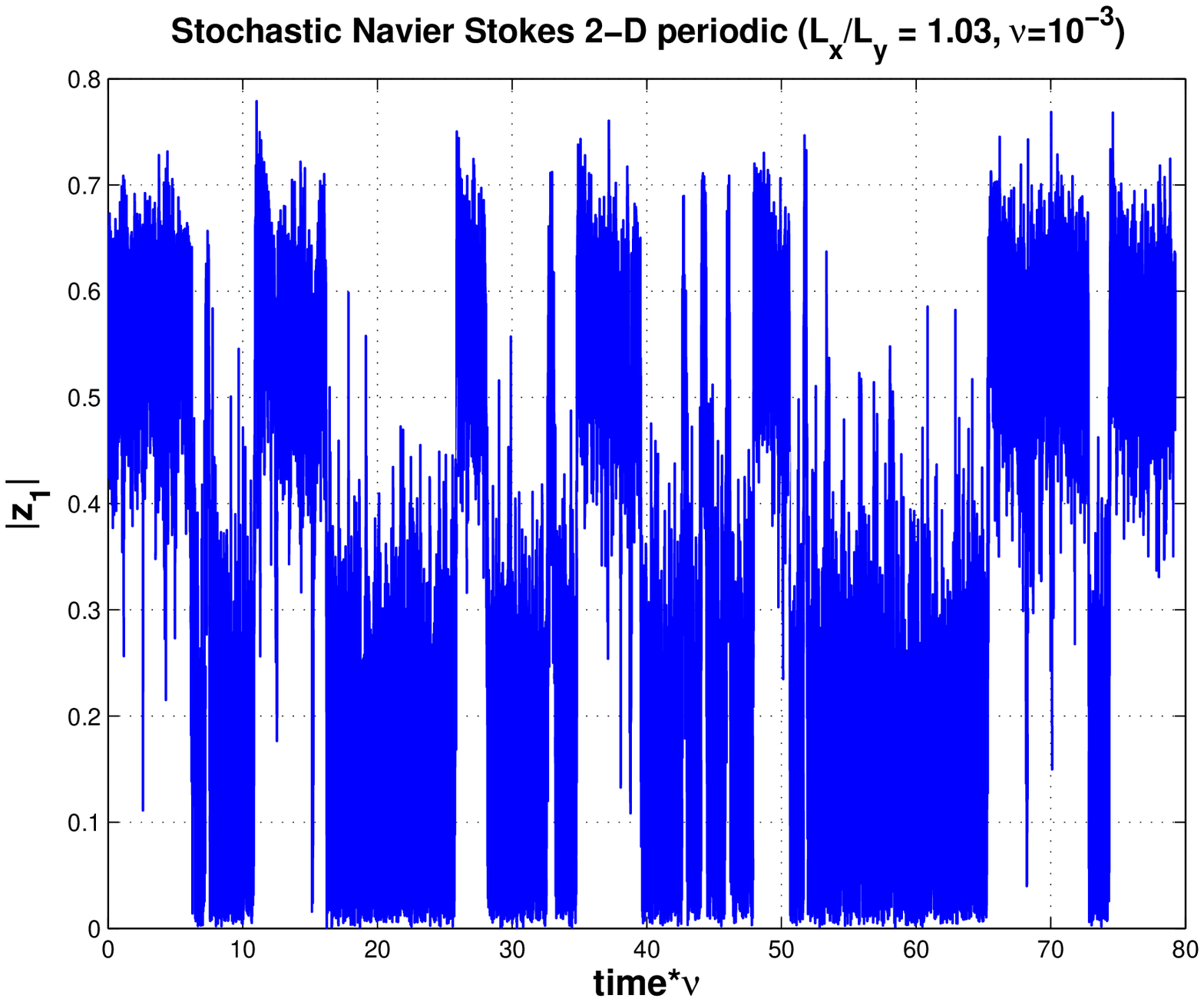}
\includegraphics[height=0.2\textheight,keepaspectratio]{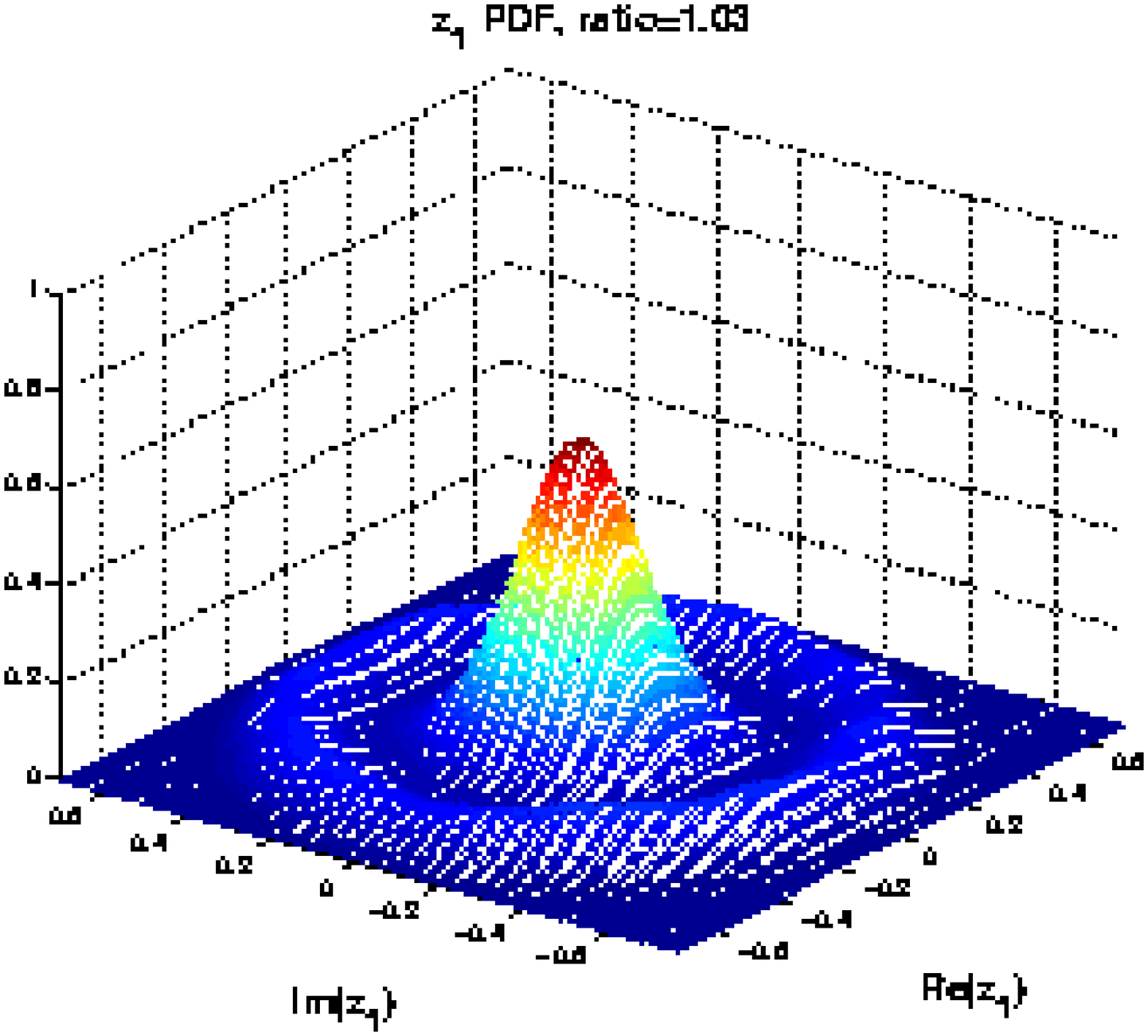}

\caption{Right panel: $\left|z_{1}\right|$, the modulus of the first Fourier
mode of the vorticity field in the direction parallel to the unidirectional
flow (for unidirectional flows, $\left|z_{1}\right|$ is close to
$0$, while for dipoles it oscillates around $0.5$. The flow thus
switches randomly from dipole to unidirectional flows. Left panel:
PDF of the complex variable $z_{1}$. \label{fig:Phase_Transition_Z1_PDF}}
\end{figure}

As shown on figure \ref{fig:Phase_Transition_Z1_PDF}, for some values
of the control parameter (the aspect ratio of the domain), we observe
the coexistence of these two flow topologies: the system switches
back and forth, at random times, between dipole and unidirectional
flows. This phenomenology is similar to what happens when noise is
added to a bistable system. A crucial difference here, is that the
deterministic dynamics does not have two different attracting states
(there is no double well potential in this case).

These few figures show that NESS, for systems with long range interactions,
may exhibit very interesting phenomena. We hope that this will open
a large number of new fundamental works on the subject. Moreover,
in a forthcoming paper, we will discuss the application of these results
to geophysical flows.

\section{Conclusion}

\label{sec:Conclusion}

We have briefly reviewed in this contribution old and new results on
the old, but active, subject of systems with long range interactions.
We clearly acknowledge that this review is far from exhaustive and
represents our personal interests.

In conclusion, it seems to us that equilibrium statistical mechanics
of these non additive systems is very well understood: a careful
application of standard tools allows one to deal with the unusual non
additivity condition, see the section devoted to equilibrium. The
situation is somewhat similar as far as relaxation to equilibrium is
concerned: in this case also, classical tools, namely those of kinetic
theory, have proved sufficient to explain some unexpected phenomena.
Thus, we fell that there is at present no obvious need for an
alternative theory describing the relaxation of these systems with
long range interacting.

Let us note that despite the successes of these well established
theories, standard statistical mechanics at equilibrium, classical
kinetic theory concerning the relaxation to equilibrium, there remains
open questions and challenges, and room for new discoveries,
especially concerning the relaxation; we have tried to outline a few
of them along the way. The most important are probably, on one hand
the quest for ensemble inequivalence, negative specific heat and phase
transitions in natural phenomena or laboratory experiment, and on the
other hand the understanding of the limits of kinetic theories.
However, we feel that the most relevant questions, both theoretically
and practically, concern forced and dissipative systems, out of
equilibrium. The last section presents very recent preliminary steps
towards an undertanding of these situations, for which the theory is
far less developed.


\begin{theacknowledgments}
  We warmly thank P.H.~Chavanis, G.~De Ninno, D.~Fanelli, K.~Jain,
  D.~Mukamel, J.~Sommeria, T.~Tatekawa and Y.~Yamaguchi for their
  participation to some of the works described in this proceeding. We
  thank especially T.~Dauxois and S.~Ruffo for their role in rising
  our interests for systems with long range interactions, and for the
  active collaborations we had during the last years, in Firenze and
  in Lyon.

  Our thought also go to Dieter Gross, the partner of so many
  discussions, and whom we remember fondly and vividly.

  This work was supported by the ANR program STATFLOW
  (ANR-06-JCJC-0037-01).
\end{theacknowledgments}



\bibliographystyle{aipproc}   


\bibliography{FBouchet,Long_Range,Meca_Stat_Euler,Experimental_2D_Flows,Euler_Stability}


\end{document}